\documentclass[reprint,amsmath,amssymb,aps,prapplied,floatfix]{revtex4-2}
\makeatletter
\usepackage{listings}
\usepackage[hidelinks,colorlinks=true,linkcolor=blue,citecolor=blue, urlcolor = blue]{hyperref}
\usepackage{bookmark}
\newcommand*{\rom}[1]{\expandafter\@slowromancap\romannumeral #1@}
\usepackage{parskip}
\usepackage{graphicx}
\usepackage{dcolumn}
\usepackage{bm}
\usepackage[utf8]{inputenc}

\begin{document}
\preprint{APS/123-QED}

\title{Dual Parton Model for the Charged Multiplicity in p-p Collisions at 13, 13.6 TeV and for future LHC energy of 27 TeV}
\author{Pranay K.~Damuka} 
\email{pranaydamuka546@gmail.com}
\affiliation{Department of Technology, Savitribai Phule Pune University, Pune 411007, India}

\author{R.~Aggarwal}
\email{ritu.aggarwal1@gmail.com}
\affiliation{Department of Technology, Savitribai Phule Pune University, Pune 411007, India}

\author{M.~Kaur}
\email{manjit@pu.ac.in}
\affiliation{Department of Physics, Panjab University, Chandigarh 160014, India\\
Department of Physics, Amity University, Punjab, Mohali {140306}, India}
\date{\today}
\begin{abstract}
Analysis of the charged multiplicity in proton-proton inelastic interactions at the LHC energies in the setting of Dual Parton Model is presented.~Data from the CMS experiment and the data simulated at different energies in various pseudo-rapidity windows using the event generator PYTHIA8 are analysed and compared with the calculations from the model.~Each inelastic scattering is assumed to follow the Poisson distribution.~The theoretical Koba-Nielsen-Olesen (KNO) scaling of the multiplicity distributions is studied and compared with previously published experimental results at $\sqrt{s}$ = 0.9, 2.36, 7 TeV.~Predictions from the model for the KNO distributions at $\sqrt{s}$ = 13, 13.6~TeV and for the future LHC energy of 27 TeV are computed and compared with the simulated data.
\end{abstract}
\maketitle
\section{\label{sec:level1}INTRODUCTION}
The number of charged particles produced in a high energy particle interaction gives a measure of an important quantity known as multiplicity.~One can validate the predictions from a theoretical or phenomenological model by studying the multiplicity distribution~(MD) of the charged particles obtained from the experimental observations.
Interest in MD was stimulated by a paper of Koba, Nielson and Olesen \cite{kno} in 1972 in which they established a scaling behaviour of the MD.~They showed that the multiplicity follows a universal scaling behaviour at high energies and termed it as the KNO scaling. 
The scaled distribution, with the parameters $z=n/\langle{n}\rangle$ and $\Psi(z)= \langle{n}\rangle P_{n}$ is independent of energy, where, $P_{n}$ is the probability of producing $n$ charged hadrons with a mean value of $\langle{n}\rangle$.~The very first KNO violation was observed at the CERN collider in $\overline{p}p$ collisions at $\sqrt{s}$ = 540~GeV~\cite{ref:knoVio}.~Soon after, the violation was also reported by different experiments involving  particle interactions at other energies with  different species of particles ~\citep{ref:knoVio1, ref:knoVio2, ref:knoVio3, ref:knoVio4}. 
In the present work, analysis of the multiplicity distributions in $pp$ interactions at the Large Hadron Collider (LHC) energies is described in the framework of the Dual Parton Model~(DPM)~\citep{236,c23,c23_1,c23_2}.~The use of DPM to describe the soft interactions started with a single Pomeron exchange.~Later, multiple Pomeron exchange diagrams were included and referred to as multiple scattering DPM.~The DPM makes use of the dual topological unitarization~(DTU) scheme~\citep{DTU1,DTU2a, DTU2b,DTU3, DTU4} to make unitarity cuts on the Pomeron exchange diagrams.~In this model a weight is associated with the cross section of each diagram appearing in the DTU expansion.~The use of DPM to define the hadron multiplicity is described briefly in the Section~\ref{sec:Theory}.

The DPM has been observed to describe the experimental charged particle multiplicities not only at ISR energies but could also account for scaling violation observed at the UA5 experiment~\citep{c23,c23_1}.~The success of the DPM is limited not only to the hadron-hadron collisions, but it has been extended to the hadron-nucleus~\citep{capellapN,capellapN_1,capellapN_2,capellapN2} and nucleus-nucleus~\citep{capellaNN, chao,capellapN_1} collisions as well.~The charged particle multiplicities data from the non-single diffractive~(NSD) $pp$ and $AA$ collisions at the LHC energy of 2.76~TeV have also been found to follow the predictions for non-diffractive~(ND) charged multiplicities calculated from the DPM~\citep{capella,72}.~The contribution from the double diffractive~(DD) processes is small~\citep{NSD,NSD2} which makes the NSD experimental data and ND calculations from the model comparable.

The ability of the DPM to describe the hadronic spectra from high energy collisions of particles of various species  motivates us to study the $pp$ collisions in the framework of the DPM.~The LHC has started producing data for $pp$ collisions at $\sqrt{s}$ = 13.6~TeV, which is the highest energy that has been ever achieved.~In this paper, the experimental charged particle multiplicities from the CMS collaboration at $\sqrt{s}$ = 0.9, 2.36, 7~TeV~\citep{cms} are  studied using the DPM.~A comparison of the charged particle multiplicities from the DPM to the experimental values is also presented in the KNO form and are found to be in agreement within the experimental limits. The average charged multiplicities for these collisions are calculated using DPM and are compared to the experimental results.~For each of the collision energies, simulated data is produced using PYTHIA 8 and compared to the predictions from the DPM and to the results from the CMS experiment. 

 The main highlights of this paper are the predictions of the charged particle multiplicities, KNO distributions and average multiplicities for the present LHC RUN3 energy of 13.6~TeV and the future LHC energy of 27~TeV in addition to the LHC RUN2 energy of 13~TeV using the DPM. These calculations are presented for the inelastic $pp$ scatterings in two central pseudo-rapidity intervals of 0.5 and 2.4. Simulated data are produced for the NSD $pp$ collisions at the energies of 13, 13.6 and 27~TeV in the two psuedo-rapidity intervals.~A detailed comparison of the KNO distributions and average multiplicities is presented in this paper using DPM calculations and predictions from PYTHIA 8.
\section{\label{sec:Theory}Theory of the Model}
In the DPM, an inelastic scattering is  considered as an exchange of a Pomeron between colliding hadrons, which results in two strings when a unitarity cut is made on the Pomeron.~In a multiple scattering DPM, multiple Pomerons are exchanged which give rise to twice the number of strings when DTU expansion scheme is used.~The weights, $\sigma_k$ with which different multiple scattering amplitudes combine in the multiple scattering DPM can be calculated using the eikonal model.~The weights $\sigma_k$ are proportional to the probability of observing $k$ inelastic collisions at given collision  energy $\sqrt{s}$ and are written below~\citep{72, sigmak, alphaP}.
\begin{equation}
   \sigma_{\textit{k}}(\xi) = \frac{\sigma_{P}}{\textit{k}Z}\left[1-e^{-Z}\sum_{i = 0}^{\textit{k}-1} \frac{Z^{i}}{i!}\right], k\geq1
\end{equation}
where $\xi = ln(\frac{\textit{s}}{\textit{s}_{0}}$), with $s_{0}$ = 1 GeV$^{2}$,  $\sigma_{P}= 8\pi\gamma_{P}e^{\Delta\xi}$, $Z = \frac{2C_{E}\gamma_{P}e^{\Delta\xi}}{R^{2}+\alpha_{P}^{'}\xi}$. Here $\sigma_{P}$ is the Born term given by Pomeron exchange with intercept $\alpha_{P}(0) = 1 + \Delta$. The values of other parameters are given below and are taken from~\citep{72}\\
$R^{2}=3.3$ GeV$^{-2}$, $\gamma_{P}=0.85$  GeV$^{-2}$, $C_{E} = 1.8$, $\Delta = 0.19$, $\alpha_{P}'=0.25$ GeV$^{-2}$. 

One of the most important parameters in Equation (1) is $\Delta$ appearing in  the Pomeron intercept, which is taken to be 0.19 and is motivated from the studies done on the LHC $pp$ data in~\citep{72} and $\gamma^{*}p$ data in \citep{delta}.~The rise in the single particle inclusive cross section per unit pseudo-rapidity with energy is governed by $\Delta$.~Parameters $R$ and $\alpha_{P}'$ control the $t$-dependence of the elastic peak~\citep{alphaP,gamma}.~The parameter $\gamma_{P}$ is determined from the absolute normalization of the total ND inelastic cross section.~A higher value of $C_{E} = 1.8$ is needed to include the high-mass diffraction states.

From the AGK cancellation \citep{18a, 18b}, one writes
\begin{equation}
\sigma_{P}(\xi)=\sum_{\textit{k}\ge{1}}\textit{k}\sigma_{\textit{k}}(\xi). 
\end{equation}

The non-diffractive inelastic scattering cross section, ${\sigma_{\textit{ND}}^{\textit{pp}}}$ can be obtained from the weights $\sigma_{\textit{k}}$ as 
\begin{equation}
{\sigma_{\textit{ND}}^{\textit{pp}}} = \sum_{\textit{k}\ge1}\sigma_{\textit{k}}(\xi).
\end{equation}

Ends of each chain are linked to the valence or sea quarks of the colliding hadrons.~The energy flow within the chains is governed by the parton distribution functions of the quarks in the colliding hadrons.~For computing the hadronic spectra in the inelastic collision, the fragmentation functions are required.~However, for the central rapidity region, the average hadronic multiplicities can be computed without using the fragmentation functions but using the rapidity position of each chain.

If we  know the number of 2$k$ chains in the $k$ inelastic scatterings with $\sigma_k$ combining cross-section, the underlying mechanism of the particle production can be predicted.
 The average number of inelastic collisions is denoted by $<k>$ and is calculated as 
\begin{equation}
<k> = \frac{{\sum_{k\ge{1}}k\sigma_{k}(\xi)}}{{\sum_{k\ge{1}}\sigma_{k}(\xi)}} = \frac{\sigma_{P}(\xi)}{\sigma_{ND}^{pp}}.
\end{equation}

Small clusters of hadrons are produced as a result of fragmentation of chains in the inelastic scatterings.~Any cluster would contain on an average $K$ number of charged particles.~A value of $K =$~1.4 is taken as it is found to describe well the hadronic spectra in $pp$ collisions~\citep{Kval1,Kval2}.~The probability of discovering  $n_{c}$ clusters of hadrons in $k$ inelastic collisions is assumed to be  given by the Poisson distribution \citep{capella} 

\begin{equation}
    \ P^{k}_{n_{c}} = e^{-k<n_{c}>_{0}}\frac{(k<n_{c}>_{0})^{n_{c}}}{n_{c}!}.
\end{equation}
Here, $<n_{c}>_{0}$ is the mean cluster multiplicity in a single inelastic scattering and it is related to the corresponding mean charged particle multiplicity $<n>_{0}$, as $ <n_{c}>_{0}=<n>_{0}/K $. 

The value of $<n>_{0}$ in a given central pseudo-rapidity region can be obtained by integrating  the charged multiplicity per unit pseudo-rapidity in an individual scattering, $\frac {dN_{0}^{\textit{pp}}} {d\eta}$, as shown below
  \begin{equation}
   \ <n>_{0} = \int_{\eta - \eta_{0}}^{\eta + \eta_{0}} \frac{dN_{0}^{pp}}{d\eta}d\eta \sim 2\eta_{0} \frac{dN_{0}^{pp}}{d\eta} \hspace{0.8mm} (\eta^{*} = 0) = 3 \eta_{0}
\end{equation}
  The quantity $\frac {dN_{0}^{\textit{pp}}} {d\eta} \hspace{0.8mm} (\eta^{*} = 0) = 1.5 $, is independent of the collision energy at mid-rapidity and at high energy as described in~\citep{72}.~Using $<k>$, the value of differential charged multiplicity  in a given psuedo-rapidity interval is calculated as
  
  \begin{equation}
   \frac{dN^{pp}}{d\eta}  =  <k>\frac{dN_{0}^{pp}}{d\eta}
\end{equation}

The total cluster multiplicity is then given by the equation 
  \begin{equation}
  \ P_{n_{c}} = \frac{\sum_{\textit{k}\ge{1}}\sigma_{\textit{k}}P^{k}_{n_{c}}}{\sigma_{\textit{ND}}^ {\textit{pp}}}.
\end{equation}
The values of $\frac{dN^{pp}}{d\eta}$, $\sigma^{pp}_{\textit{ND}}$  and $\sigma^{pp}_{\textit{tot}}$ from the DPM  at various collision energies reported in~\citep{72} are reproduced as shown in the Table~\ref{nonlin}.~Their values at the collision energies of 13, 13.6 (ongoing RUN3) and future LHC collision energy of 27~TeV are also calculated and shown in the given table.


\section{KNO Formalism}
The KNO form of the multiplicity distributions are written in terms of $\Psi(z)$
\begin{equation}
    \Psi(z) = \langle{n}\rangle P_{n} = \langle{n_{c}}\rangle P_{n_{c}}
\end{equation}
where, 
\begin{equation}z = n/\langle{n}\rangle = n_{c}/\langle{n_{c}}\rangle.
\end{equation}

Here, $n$ is the number of emitted charged particles and  $\langle{n}\rangle$ is it’s mean value.~The latter can be calculated from the average cluster multiplicity using the equation $\langle{n}\rangle = <k><n>_{0} = 3\eta_{0}<k>$. 
The relation between $n$ and $n_{c}$ is given by using $n_{c} = n/K$, also $\langle{n_{c}}\rangle=\langle{n}\rangle/K$.
\begin{table}[htbp!]
\caption{Cross sections of total and non-diffractive processes for $pp$ interactions and the charged particle pseudo-rapidity
densities in the central rapidity region corresponding to $\sqrt{s}$ \label{nonlin}}

\begin{tabular}{c c c c} 
\hline\hline 
$\sqrt{s}$ (GeV) & $\frac {dN^{\textit{pp}}} {d\eta}$ $(y^{*} = 0)$ & ${\sigma_{\textit{ND}}^{\textit{pp}}}$ &  ${\sigma_{\textit{tot}}^{\textit{pp}}}$\\ [0.8ex] 
\hline 
200   &   2.99   &   31.22   &   41.62\\
540   &   3.50   &   38.97   &   54.39\\
900   &   3.82   &   43.33   &   61.85\\
1800  &   4.34   &   49.64    &  72.93 \\
2360  &   4.57   &   52.24    &  77.54 \\
2760  &   4.71   &   53.77    &  80.27 \\
7000  &   5.70   &  63.33    &  97.57  \\
13000 &   6.50   &  70.15    &  110.03  \\
13600 &   6.57   &  70.66    &  110.96  \\
27000 &   7.66     &  78.67    &   125.68\\
\hline\hline
\end{tabular}
\end{table}
\section{The Data analysed}
The charged hadron multiplicity in NSD processes in $pp$ collisions has been measured by the CMS collaboration \citep{cms} at various center of mass energies, $\sqrt{s} = 0.9, 2.36$ and 7~TeV in restricted pseudo-rapidity intervals of $|\eta| < 0.5$ and $|\eta| < 2.4$.~These experimental results are available on the HEPData \citep{HEP} and have been used for the present study.~In addition, for detailed comparison, we used the Monte Carlo~(MC) event generator PYTHIA 8.306 for the generation of $10^{7}$ $pp$ NSD events at each $\sqrt{s}$ and in each pseudo-rapidity interval under study.                                    

~With the aim of providing a better description of the observables of the data, event generators have adjustable parameters to control the behaviour of the event modeling.~These observables may be sensitive to partons from initial-state radiation~(ISR) and final-state radiation~(FSR), underlying events~(UE) consisting of beam remnants, particles produced in multiple-parton interaction~(MPI) etc.~The processes of hadronization and MPI are particularly afflicted, as they involve non-perturbative QCD physics.~A good modeling of hadronization is required.~A set of parameters, which need to be adjusted to fit some aspects of the data, is referred to as a tune.~The experimental data from an experiment is often fitted to the predictions from an event generator by tuning and optimizing these parameters.~Description of PYTHIA as an event generator can be found in~\citep{pythia}.~For the present analysis, PYTHIA 8 tunes, Monash~\citep{Skands} and 4C~\citep{4ctune} are used for studying the charged hadron multiplicity distributions.~The MDs have also been simulated for $\sqrt{s}$ = 13, 13.6 and 27~TeV using PYTHIA 8.3 for comparison with the theory and for predicting the multiplicity at the future LHC energy of 27~TeV. 
\section{Results}
\subsection{Multiplicity Distributions}

\begin{figure}[htbp!]
{\includegraphics[width=0.48\textwidth]{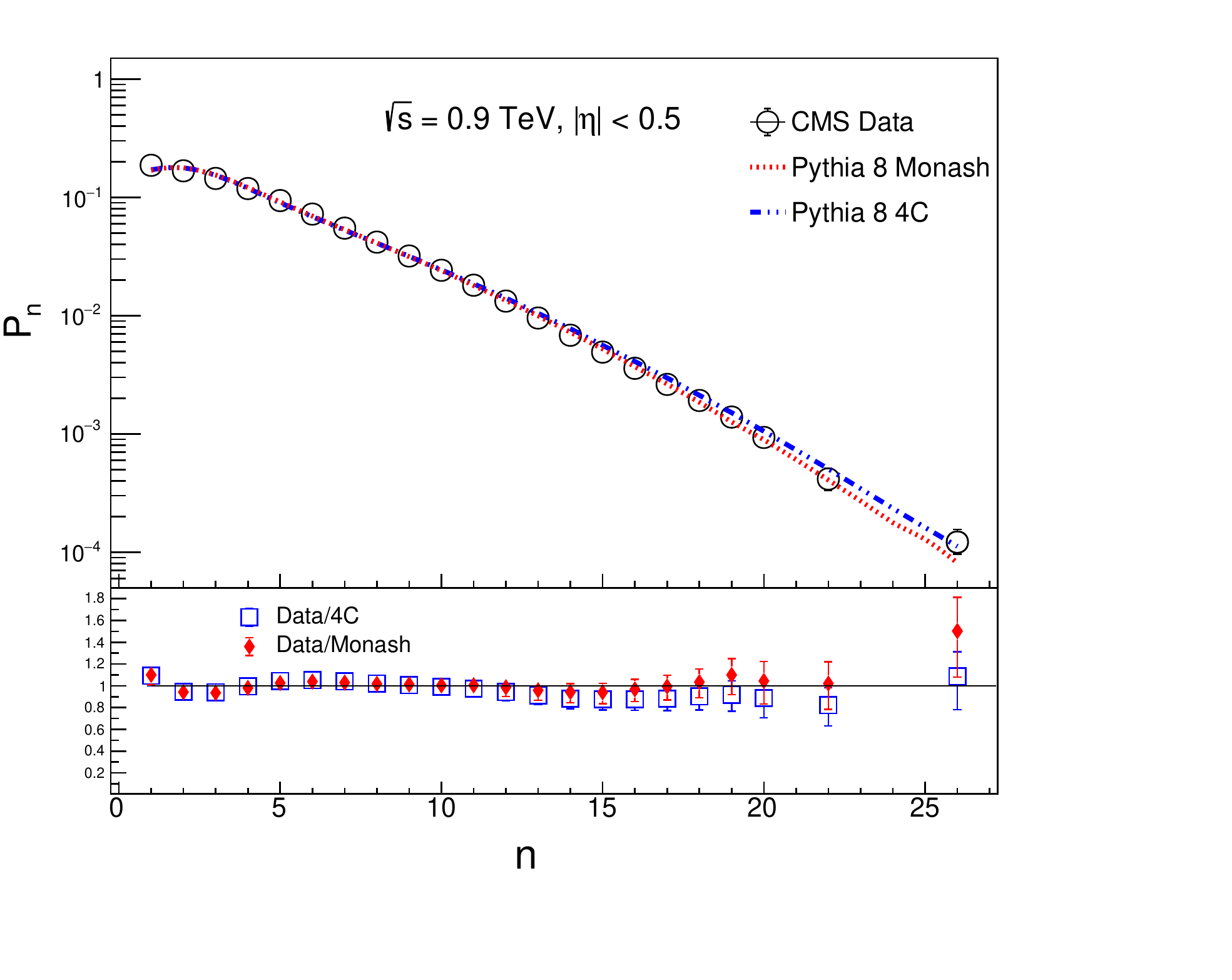}
\includegraphics[width=0.48\textwidth]{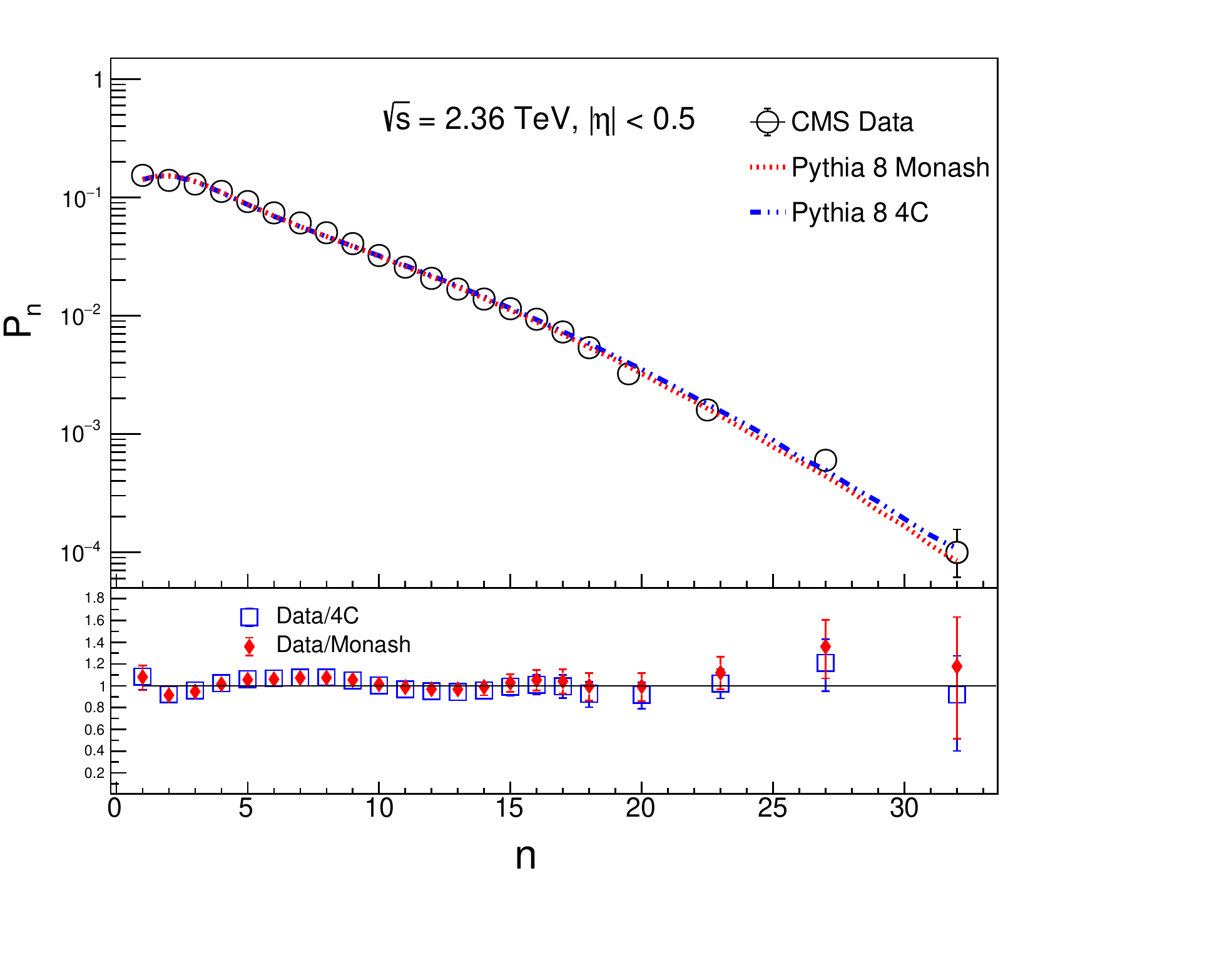}
\includegraphics[width=0.48\textwidth]{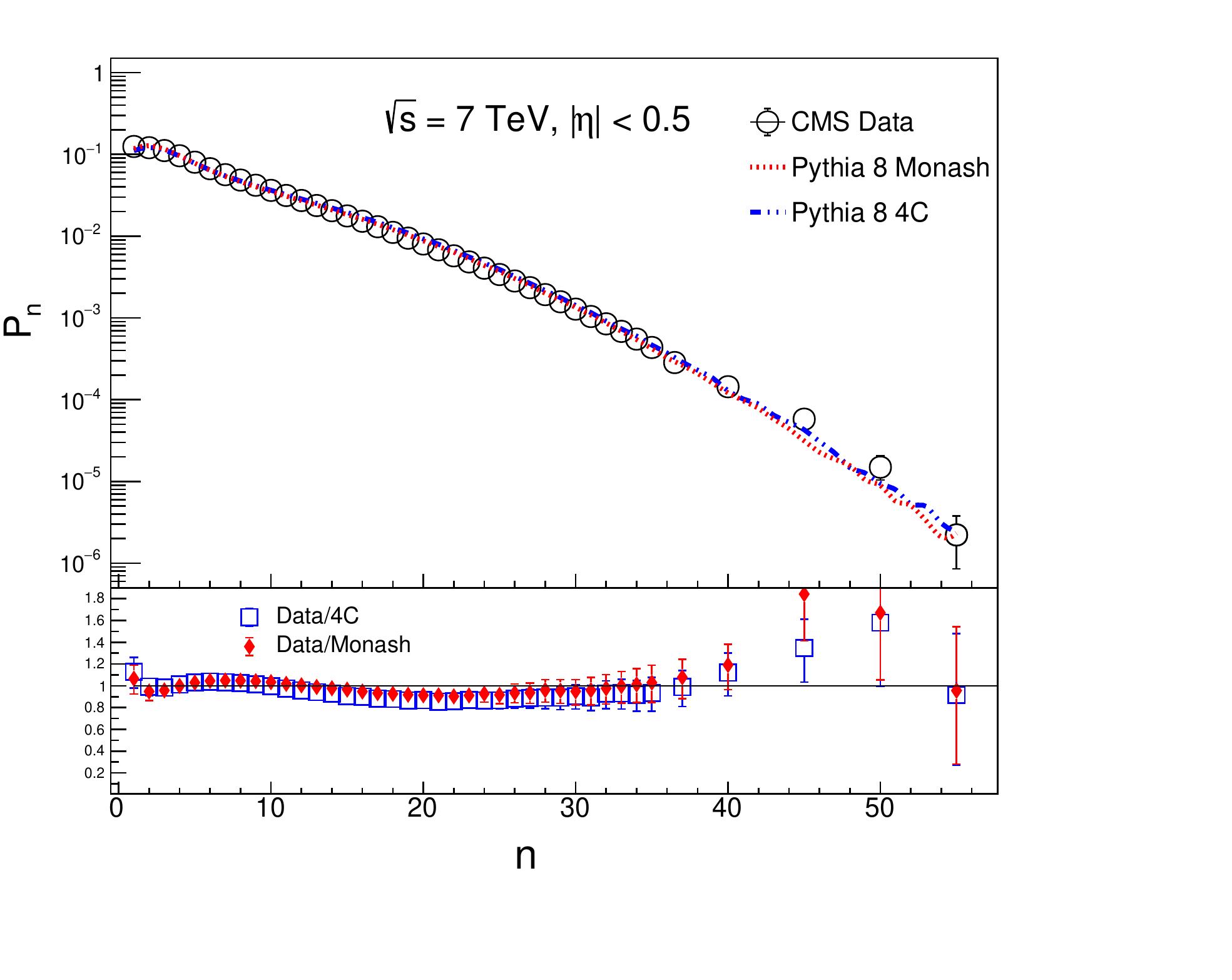}}
\caption{Charged hadron multiplicity distributions for $\sqrt{s}$ = 0.9, 2.36 and 7~TeV in the pseudorapidity $|\eta| < 0.5$.~Points represent the data obtained by the CMS collaboration and the lines represent the values from PYTHIA for two tunes.~In each plot, the lower panel shows the ratio of data versus MC tunes.} 
\label{fig:One}
\end{figure}

For a probability $P_{n}$ of producing $n$ charged particles, the mean charged multiplicity is defined as: 
\begin{equation}
    \langle{n}\rangle = \frac{\sum{nP_{n}}}{\sum{P_{n}}}
\end{equation}
Figure~\ref{fig:One} shows the charged hadron multiplicity distributions for $\sqrt{s}$ = 0.9, 2.36 and 7~TeV in the pseudo-rapidity $|\eta| < 0.5$ interval.~The data obtained by the CMS collaboration are shown and compared with the distributions simulated by using PYTHIA 8 for two different tunes, Monash and 4C.~In each plot, the lower panel shows the ratio of data versus MC tunes.
It is observed that the charged MDs agree with the experimental distributions in general, with small fluctuations at higher multiplicity tails as can be observed from the ratio plots, at each energy.

\begin{figure}[htbp!]
\includegraphics[width=0.48\textwidth]{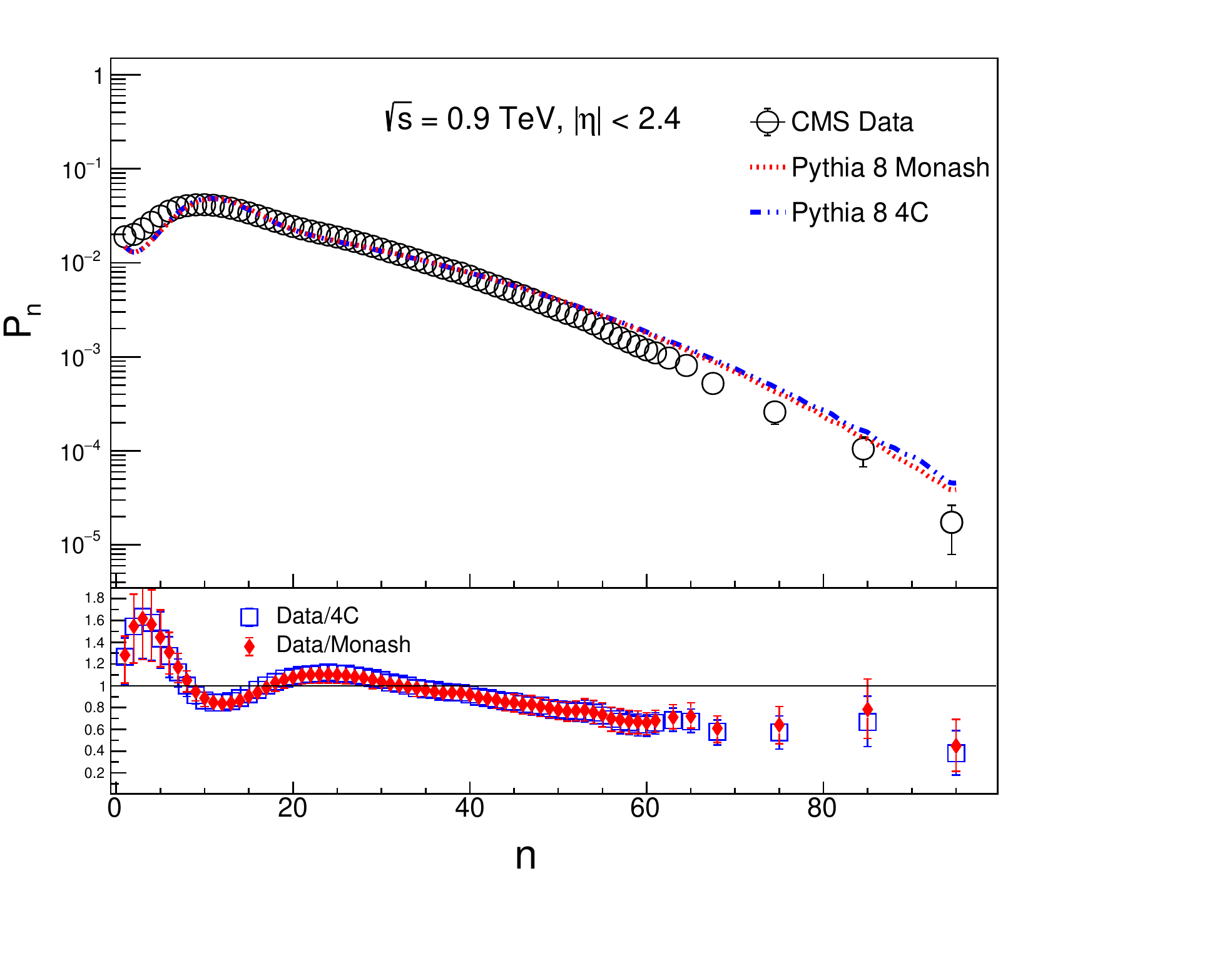}
\includegraphics[width=0.48\textwidth]{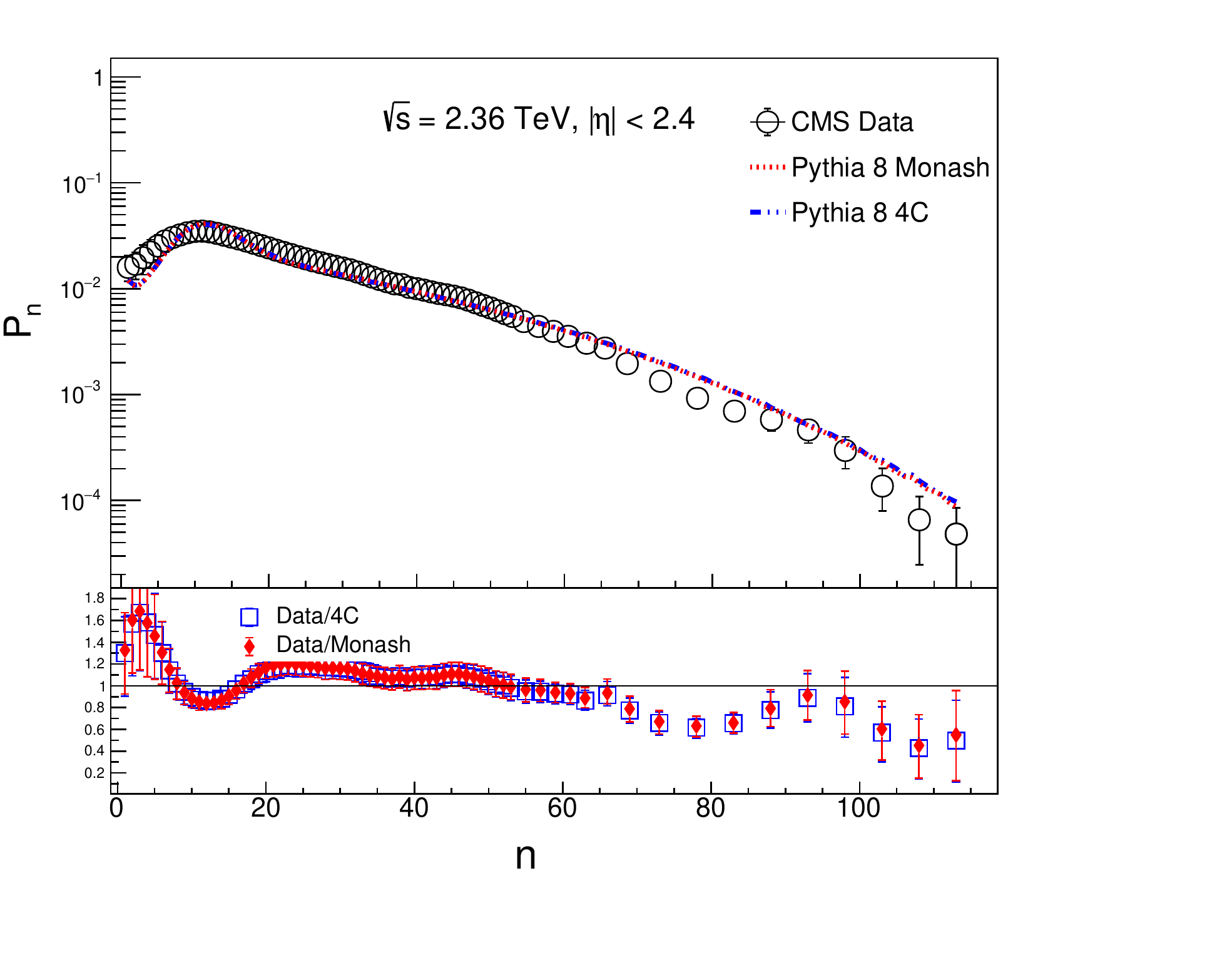}
\includegraphics[width=0.48\textwidth]{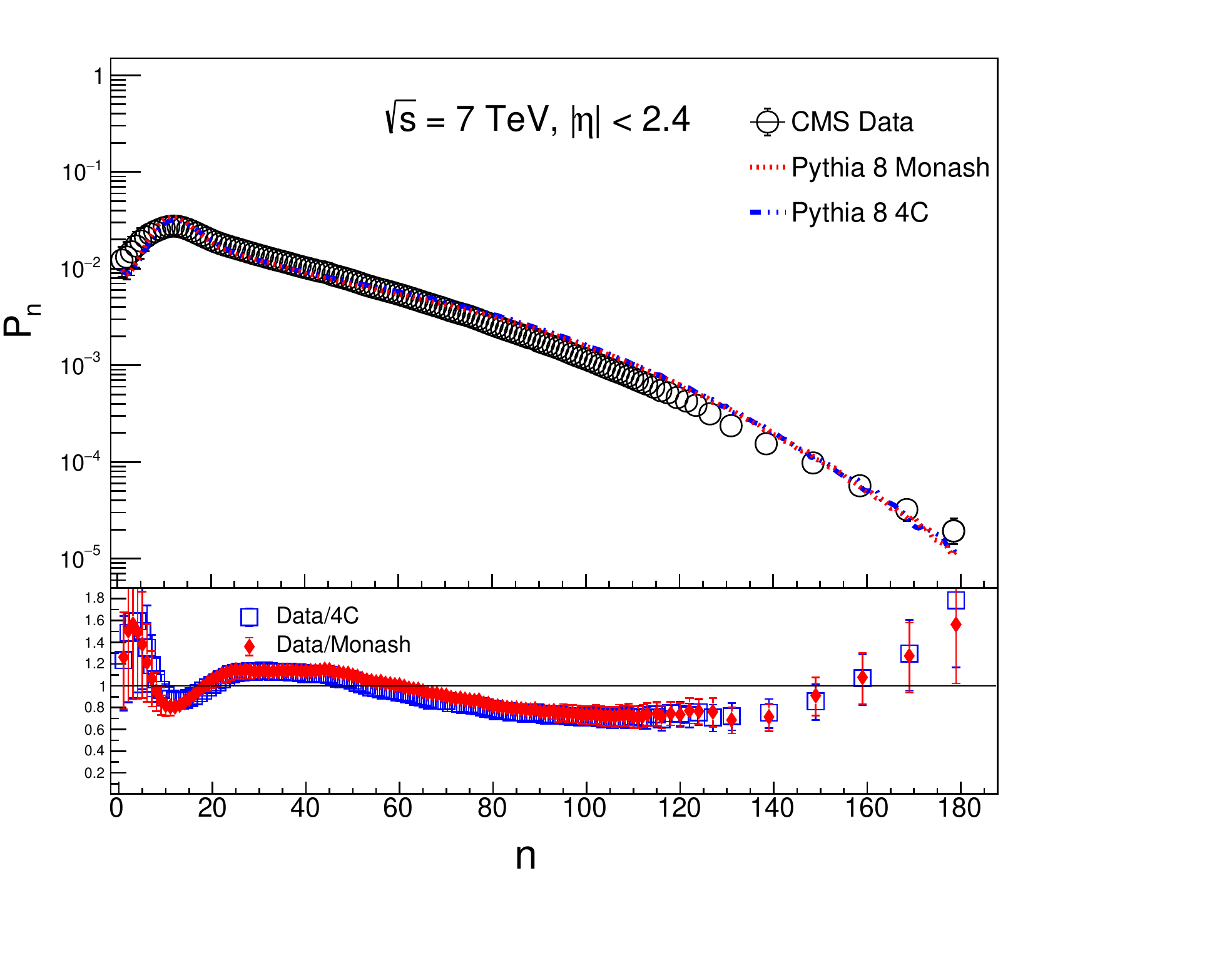}
\caption{Charged hadron multiplicity distributions for $\sqrt{s}$ = 0.9, 2.36 and 7~TeV in the pseudorapidity $|\eta| < $ 2.4 interval.~Points represent the data obtained by the CMS collaboration and the lines represent from PYTHIA for Monash and 4C tunes.~In each plot, the lower panel shows the ratio of data versus MC tunes.}
\label{fig:Two}
\end{figure}

Figure~\ref{fig:Two} shows the charged hadron multiplicity distributions for $\sqrt{s}$ = 0.9, 2.36 and 7~TeV in the pseudo-rapidity interval $|\eta| < 2.4$.~Again the data obtained by the CMS collaboration are shown and compared with the distributions simulated by using PYTHIA 8 for two different tunes, Monash and 4C.
The multiplicity distributions have different shapes in the two pseudo-rapidity intervals.~A shoulder structure in the low multiplicity region in $|\eta| <$2.4 interval can be clearly seen for all energies. 

It is observed that the ratio of the experimental MDs to the simulated data fluctuates around $\sim$ 1.~However, there is an observable deviation at mid-and-higher multiplicities.~In the mid multiplicity region, PYTHIA 8 underestimates the MDs and in the higher multiplicity region, it overestimates.~This effect gets pronounced with increase in centre of mass energy as shown in the ratio plots.~Both Monash and 4C tunes show similar behaviour at all energies.

\subsection{Mean multiplicity from DPM and PYTHIA }
\begin{figure}[htbp!]
\centering
\includegraphics[width=0.48\textwidth]{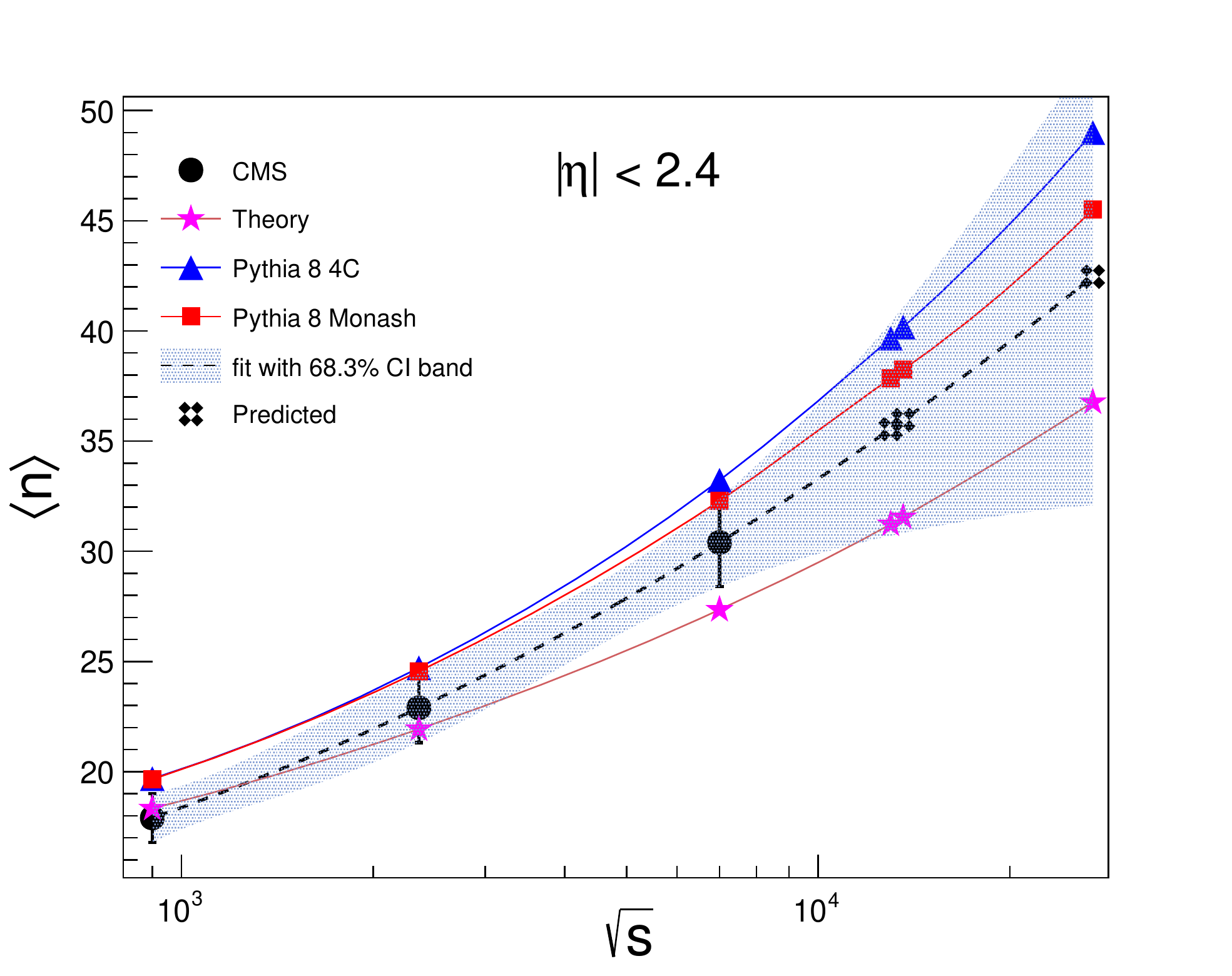}
\caption{Mean multiplicity $\langle{n}\rangle$ as a function of $\sqrt{s}$ for $|\eta|<$ 2.4}\label{fig:AvnRoots}
\end{figure}

Table~\ref{table:pyt} presents the mean multiplicity at different center of mass energies for the CMS data \citep{cms}, for DPM described in section~\ref{sec:Theory} and for PYTHIA 8 tunes, Monash and 4C in $|\eta| <2.4$ interval.~It is observed that the theoretical values from DPM and those from PYTHIA 8 agree with the experimental values within the error limits.~However the theory tends to underestimate the experimental mean multiplicity while both tunes of PYTHIA slightly overestimate.
Figure~\ref{fig:AvnRoots} shows the variation of mean multiplicity $\langle{n}\rangle$ with $\sqrt{s}$ from 0.9 to 27~TeV.~Data from the CMS experiment are shown in comparison to the theory and MC simulations from PYTHIA for the tunes Monash and 4C.~The variation in each case can be represented by a fit of the type:
\begin{equation}\label{EqRoots}
     \langle{n}\rangle = a + b\ln{\sqrt{s}}+c(\ln{\sqrt{s}})^2
\end{equation}
For the CMS data the fit parameters are given by $a = 26.70\pm 2.63, b = -13.94\pm0.76$ and $c = 3.34\pm 0.16$.
An extrapolation of the fit to higher $\sqrt{s}$ predicts the $\langle{n}\rangle$ as shown in the table~\ref{table:pyt}. The $68.3\%$ confidence interval (CI, 1$\sigma$) band on the fit function is shown in Figure~\ref{fig:AvnRoots} along with the prediction for experimental $\langle{n}\rangle$ at 13, 13.6 and 27 TeV. The values at these higher energies from the DPM and two PYTHIA tunes lie within the $68.3\%$ CI band.

\begin{table}[htbp!]
\caption{$\langle{n}\rangle$ as a function of $\sqrt{s}$ for $|\eta| < 2.4$.~For each $\sqrt{s}$ a sample of 10 million events is simulated for every $\langle{n}\rangle$.~Thus the statistical errors are negligible and hence have not been quoted.~Values shown with $(*)$ are predicted from (\ref{EqRoots})}. 
\begin{tabular}{c c c c c} 

\hline\hline 
$\sqrt{s}$ (TeV) & Experiment~\citep{cms} & Theory &  4C & Monash\\ [0.8ex] 
\hline 
0.9 & 17.9$\pm0.1_{-1.1}^{+1.1}$ & 18.33& 19.68 & 19.65\\ 
2.36 & 22.9$\pm0.5_{-1.5}^{+1.6}$ & 21.93& 24.71 & 24.55\\
7 & 30.4$\pm0.2_{-2.0}^{+2.2}$ & 27.35& 33.22 & 32.29\\
13 &  35.60$^{(*)}$  &  31.23    & 39.67 & 37.85\\
13.6 & 36.00$^{(*)}$ &  31.55    & 40.17 & 38.24\\
27 &  42.52$^{(*)}$  &  36.77    & 49.00 & 45.51\\
\hline\hline
\end{tabular}
\label{table:pyt} 
\end{table}
 
\subsection{KNO Multiplicity Distributions}
\begin{figure}[htbp!]
\centering
\includegraphics[width=0.48\textwidth]{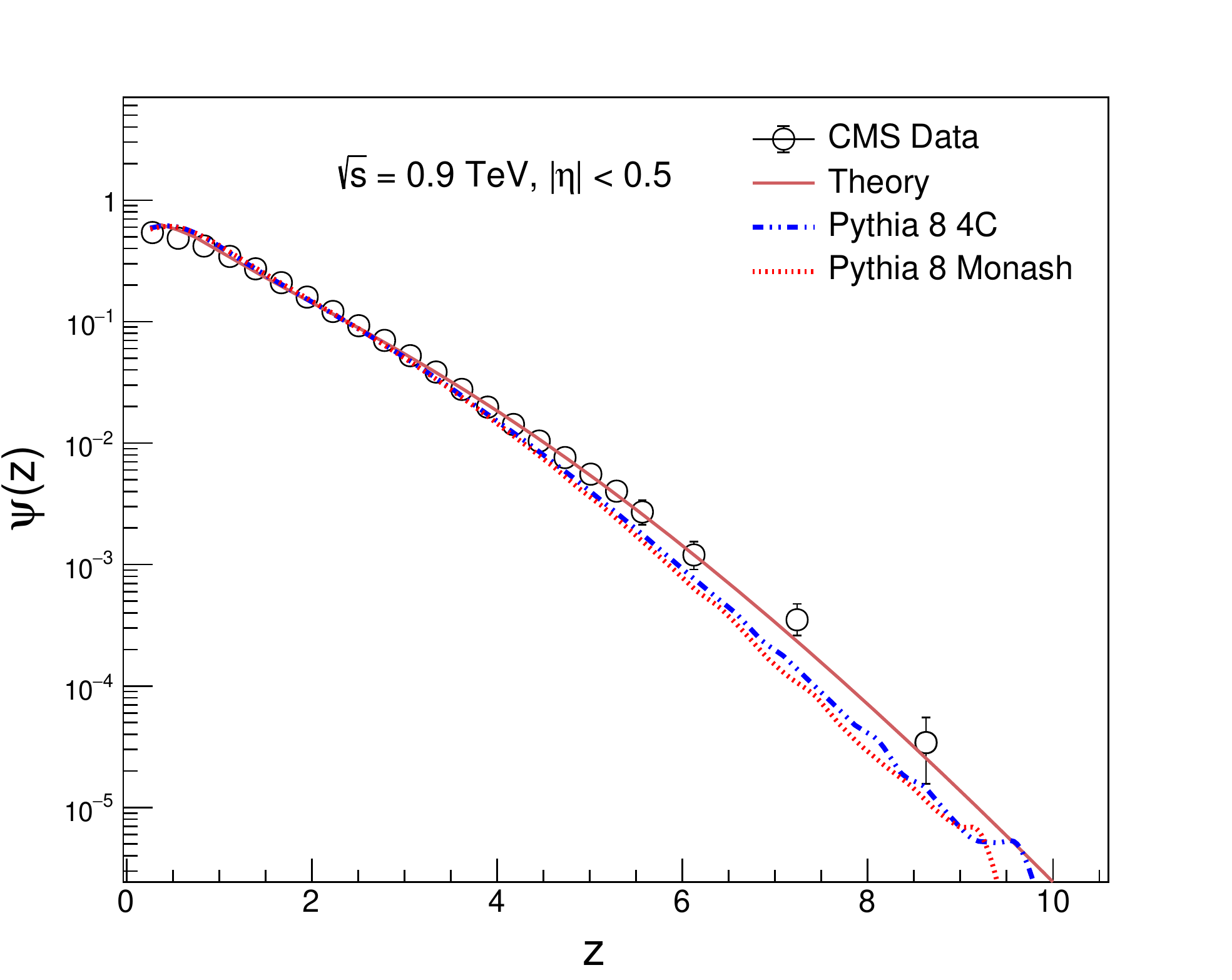}
\includegraphics[width=0.48\textwidth]{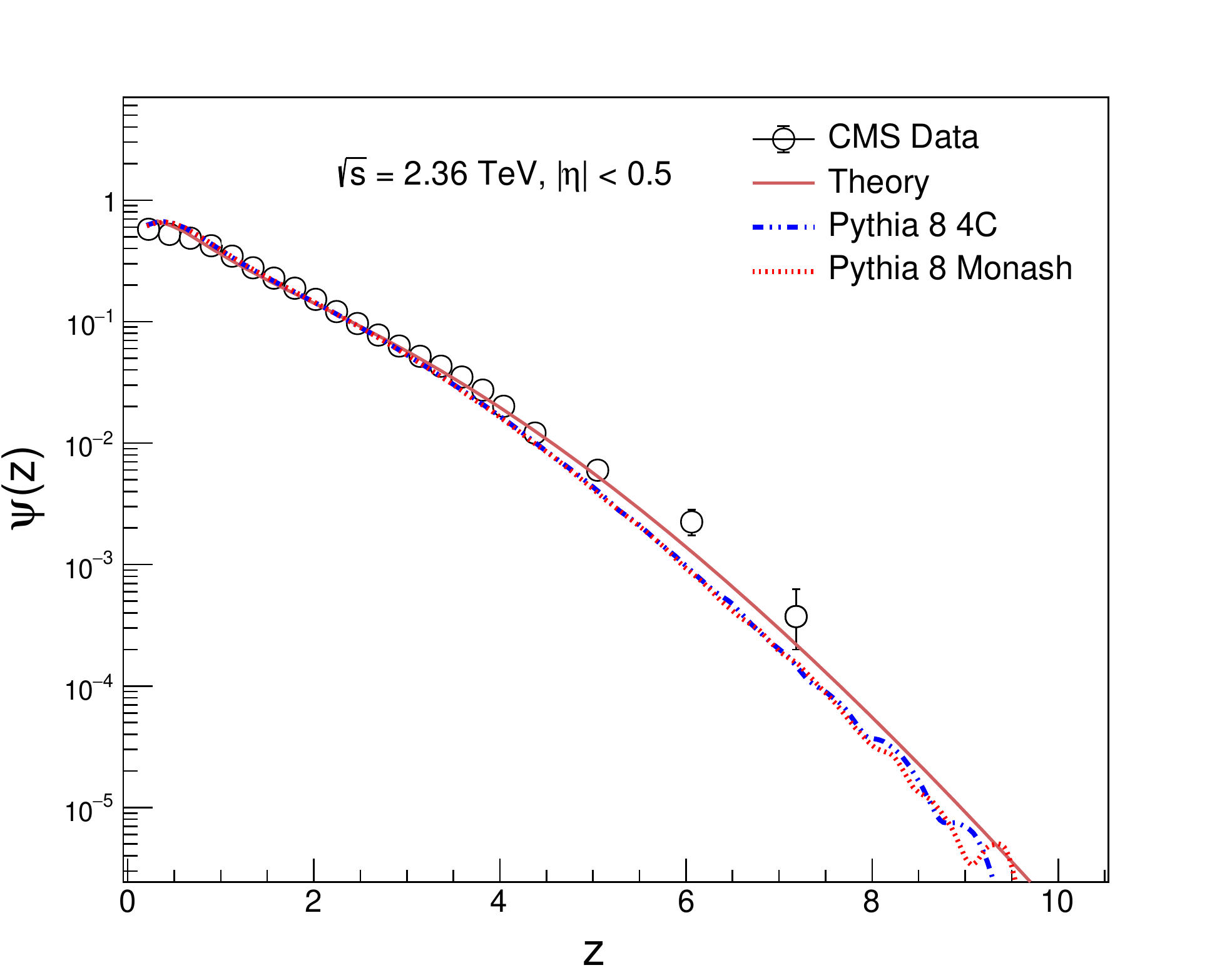}
\includegraphics[width=0.48\textwidth]{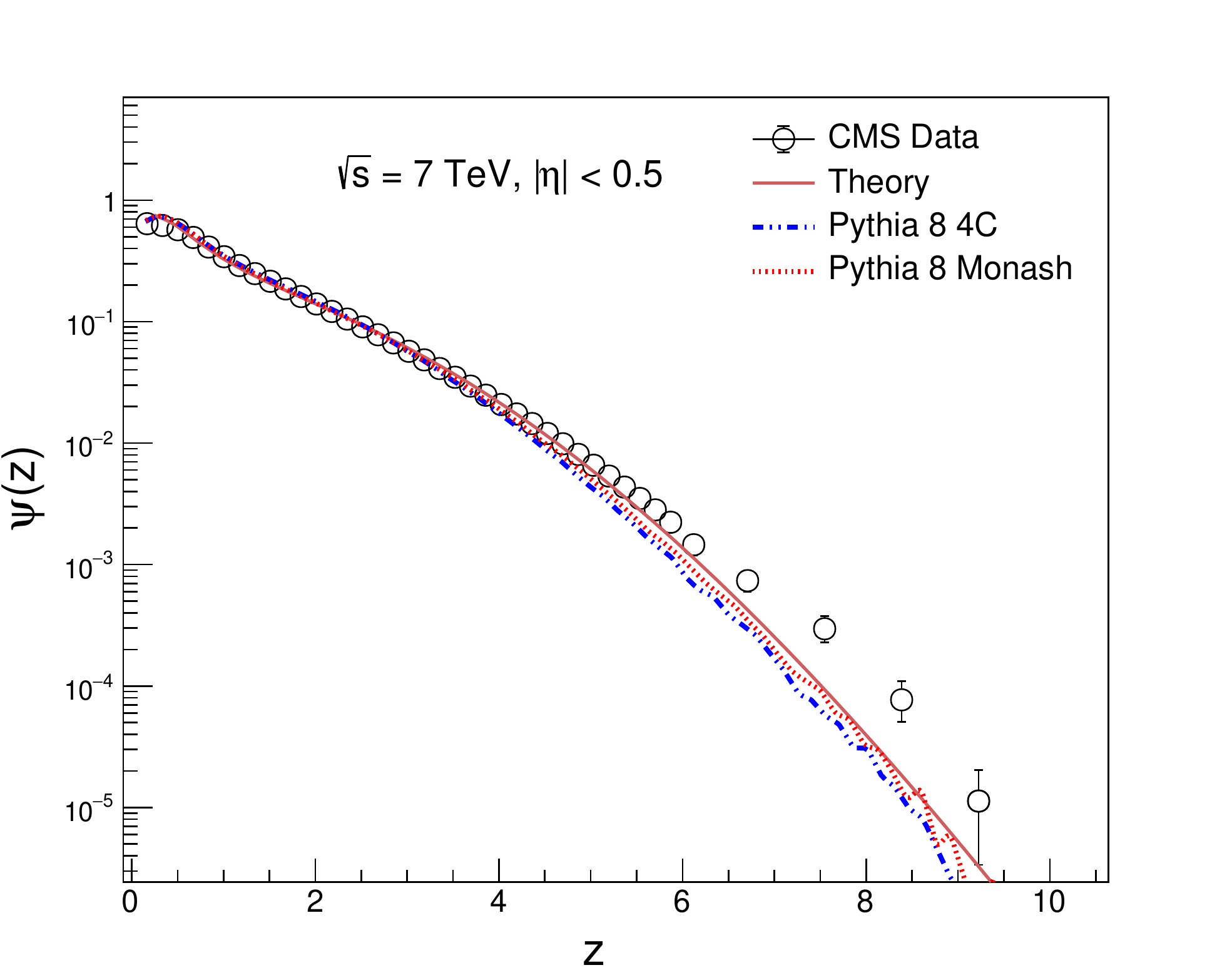}
\caption{KNO distributions at $\sqrt{s}$ = 0.9, 2.36 and 7~TeV in the pseudorapidity $|\eta| < 0.5$.~Points represent the data obtained by the CMS experiment, solid lines represent the predictions from the theoretical model and dotted lines represent the distributions generated from PYTHIA for two different tunes.}
\label{fig:Kno3}
\end{figure}

\begin{figure}[htbp!]
\centering
{\includegraphics[width=0.48\textwidth]{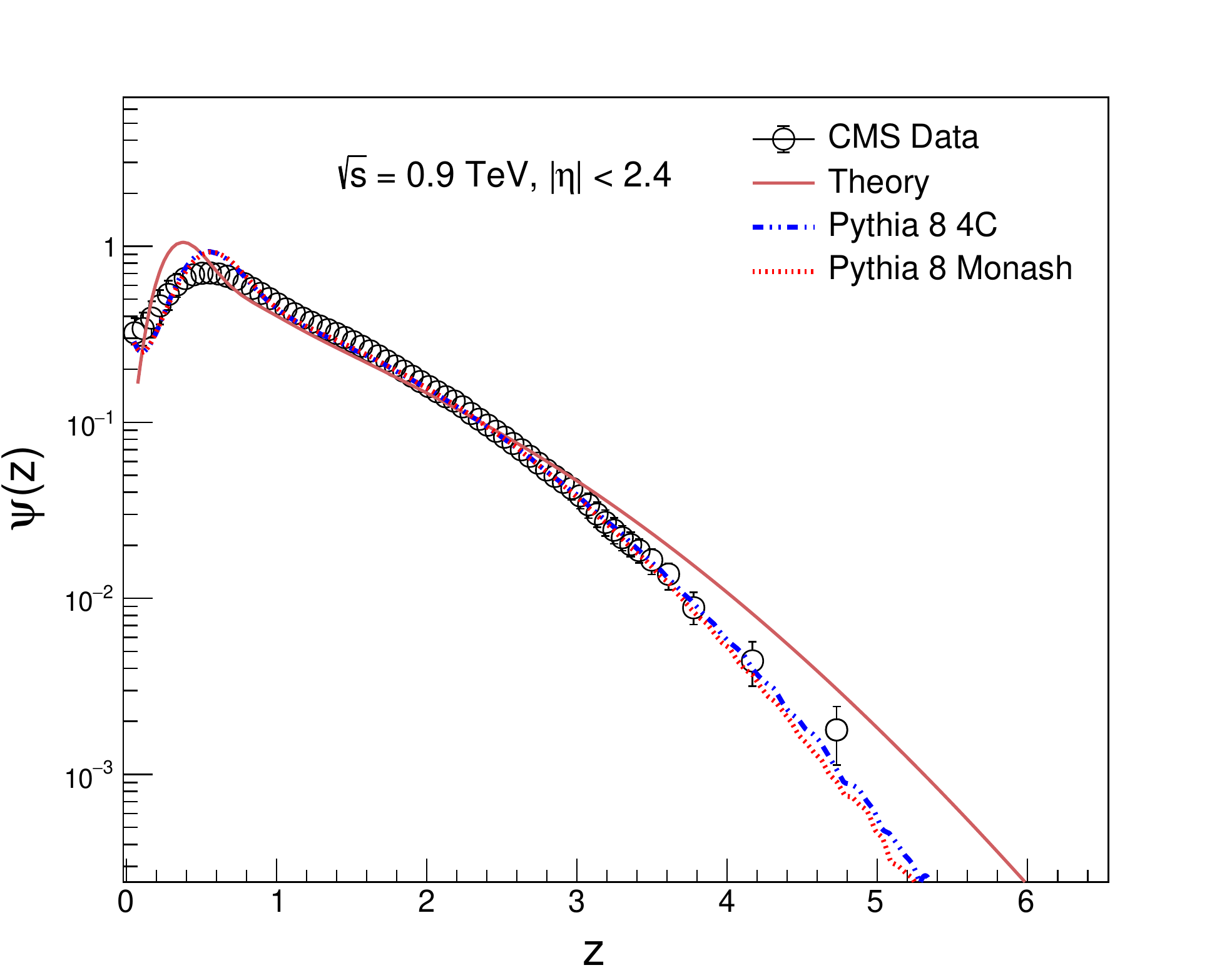}
\includegraphics[width=0.48\textwidth]{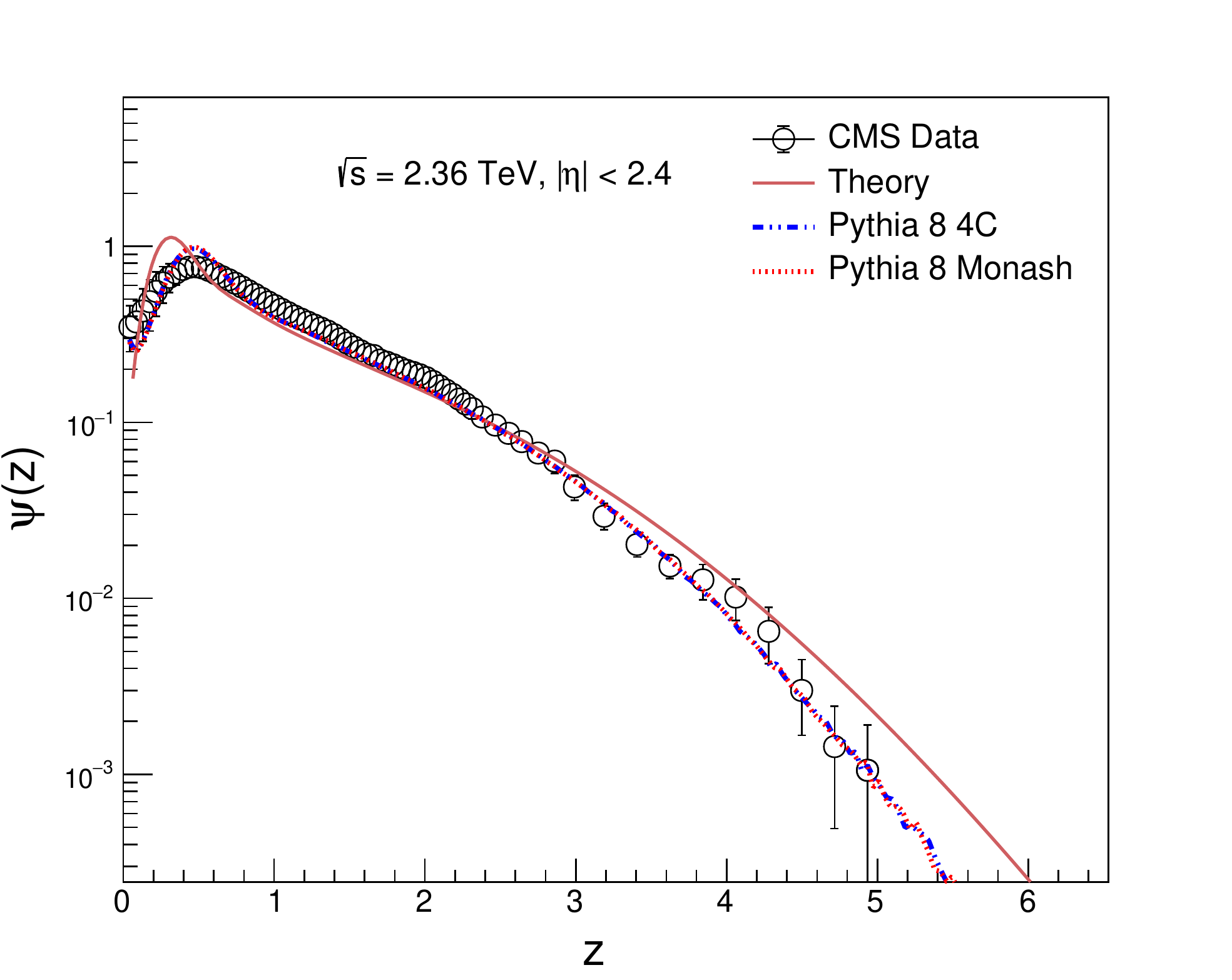}
\includegraphics[width=0.48\textwidth]{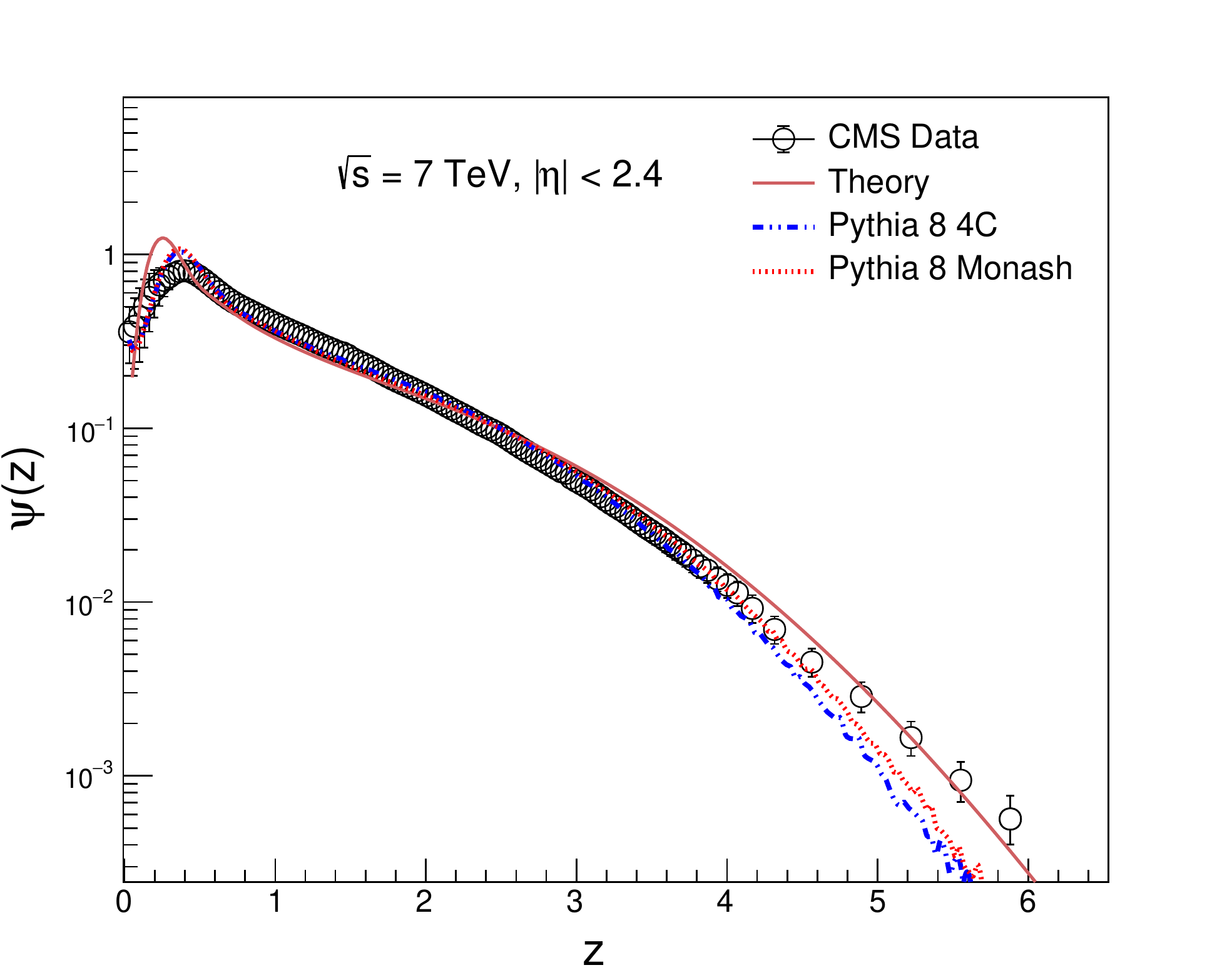}}
\caption{KNO distributions at $\sqrt{s}$ = 0.9, 2.36 and 7~TeV in the pseudorapidity $|\eta| < 2.4$.~Points represent the data obtained by the CMS experiment, solid lines represent the predictions from the theoretical model and dotted lines represent the distributions generated from PYTHIA for two different tunes.}
\label{fig:Kno4}
\end{figure}

The KNO multiplicity distributions in $\Psi(z)$ versus $z$ are presented in figures~\ref{fig:Kno3} and \ref{fig:Kno4} for various centre of mass energies in $|\eta| <$ 0.5 and $|\eta| <$ 2.4 intervals respectively.~The experimental data \citep{HEP} are compared with the theoretical predictions from the DPM and with the MC data simulated from PYTHIA for the tunes, Monash and 4C.~For $|\eta| <$ 0.5, it is observed that the PYTHIA tunes agree with the data at lower $z$, but show increasing disagreement above $z\sim$ 4 at all energies.~The deviation of the KNO distributions from the DPM starts at $z\sim6$ for $|\eta| <$ 0.5 at all energies.~For $|\eta| <$ 2.4 interval, the agreement with both PYTHIA tunes is better as seen in figure~\ref{fig:Kno4} and the deviation is observed only at 7~TeV for $z>$ 4.~However, in this region the model deviates from the data at all $\sqrt{s}$ for $z > $ 3.

A shoulder structure is observed for each distribution in the $|\eta| <$ 2.4 region.~The model is not able to describe the shoulder structure fully for all energies.~The data when compared with MC distributions from PYTHIA show an agreement better than the model, and the agreement improves with the increasing collision energy.~For a better comparison a blown-out view of every distribution in the region around peak is presented for the higher pseudo-rapidity region.

\begin{figure}[htbp!]
\centering
\includegraphics[width=0.48\textwidth]{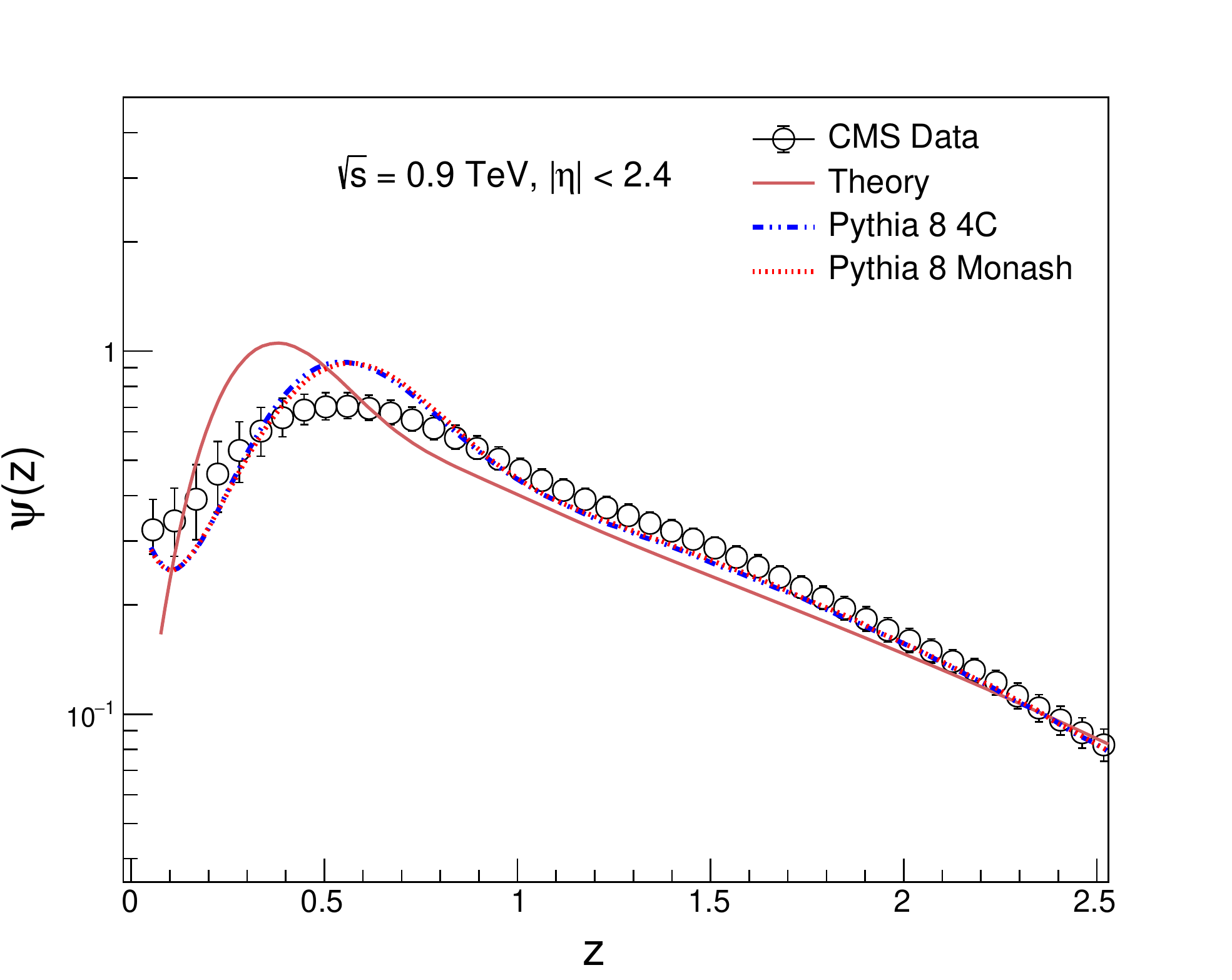}
\includegraphics[width=0.48\textwidth]{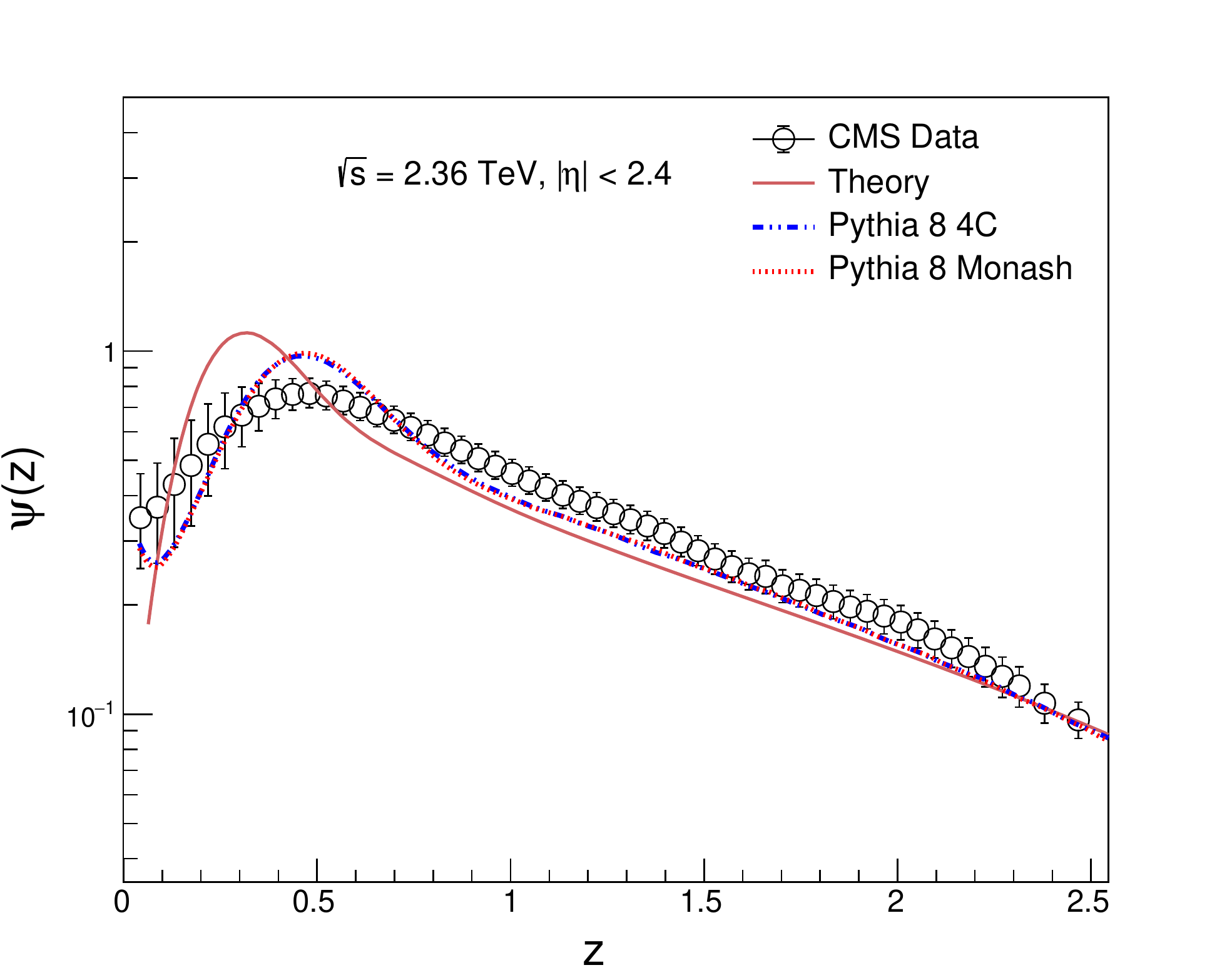}
\includegraphics[width=0.48\textwidth]{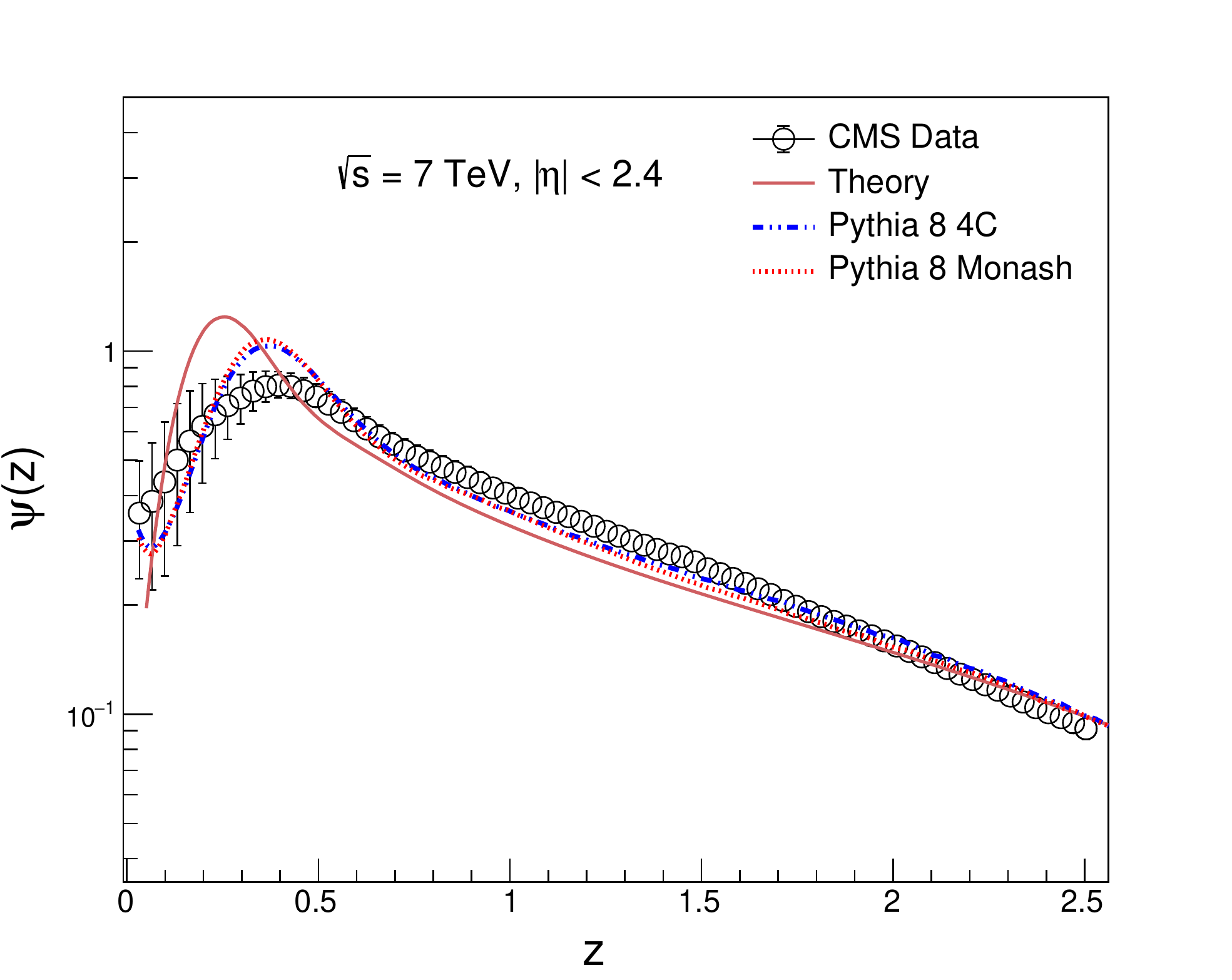}
\caption{KNO distributions in the range $z$ = 0 to 2.5 for different c.m.s energies, in the pseudorapidity interval $|\eta| < 2.4$.~Solid lines represent the predictions from the theoretical model and dotted lines represent the distributions generated from PYTHIA for two different tunes at each energy.}
\label{fig:peak1}
\end{figure}

Figure~\ref{fig:peak1} shows the KNO distributions around the peak values for $z <$ 2.5 at $\sqrt{s}$ = 0.9, 2.36 and 7~TeV in the $|\eta| <$ 2.4 range. It is observed that there is a disagreement between the model predictions and the data.~Though the shoulder structure present in the data is reproduced by the model, the position of the peak is shifted to the lower multiplicities in comparison to the data.~The MC and the data distributions are peaked nearly at the same positions.~The distributions from the tunes Monash and 4C agree closely with each other and also with the data within the limits of experimental errors.

\begin{figure}[htbp!]
\centering
\includegraphics[width=0.48\textwidth]{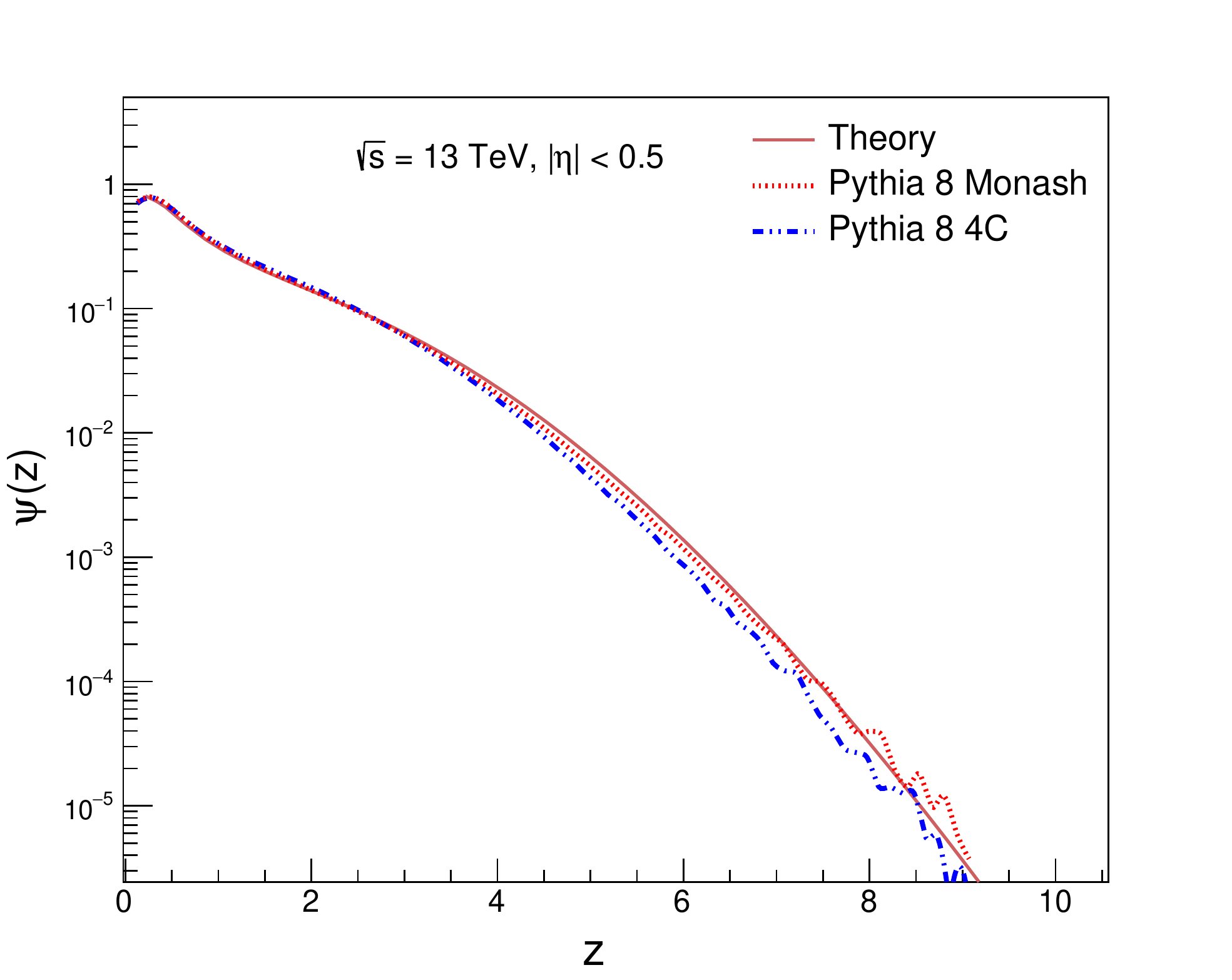}
\includegraphics[width=0.48\textwidth]{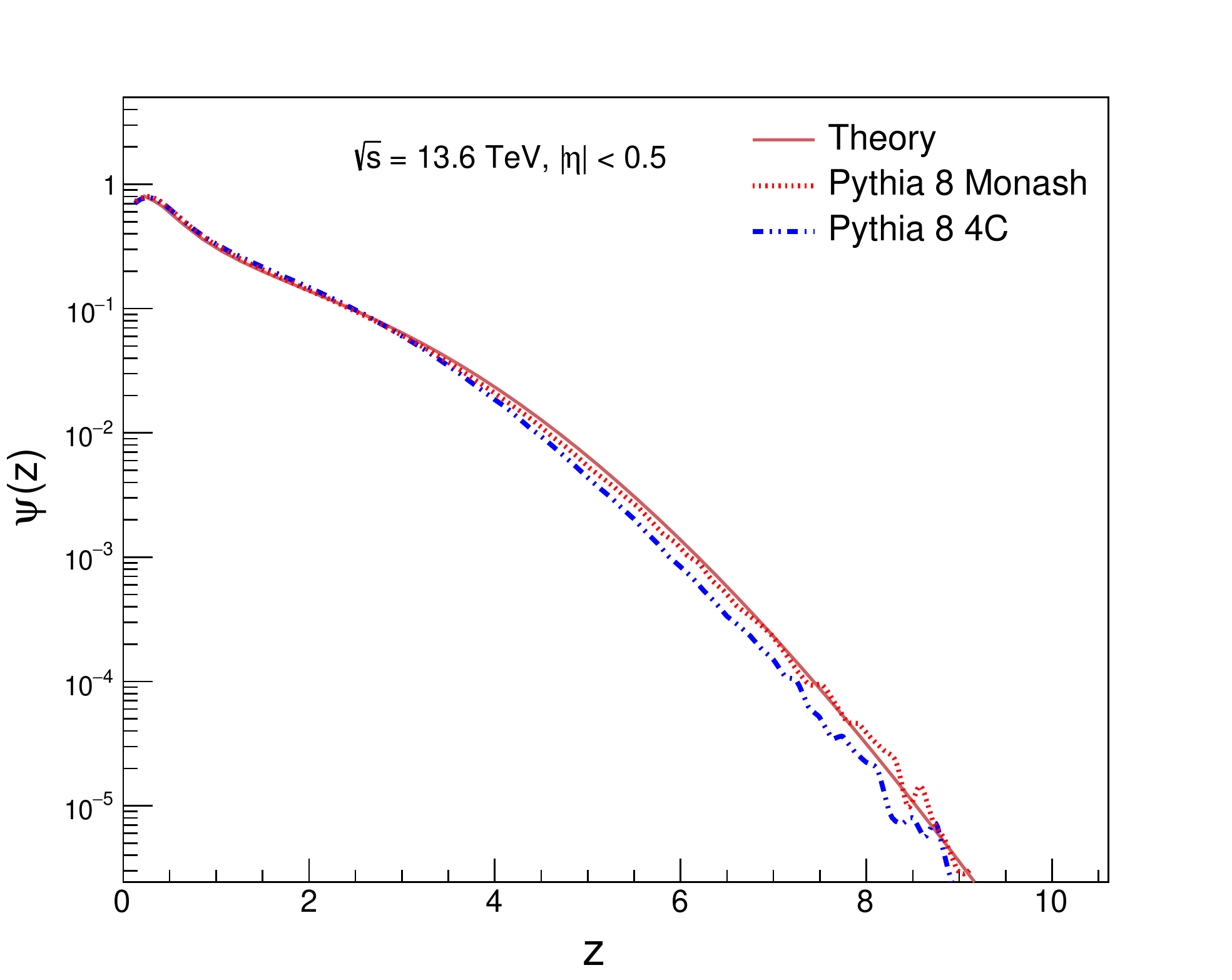}
\includegraphics[width=0.48\textwidth]{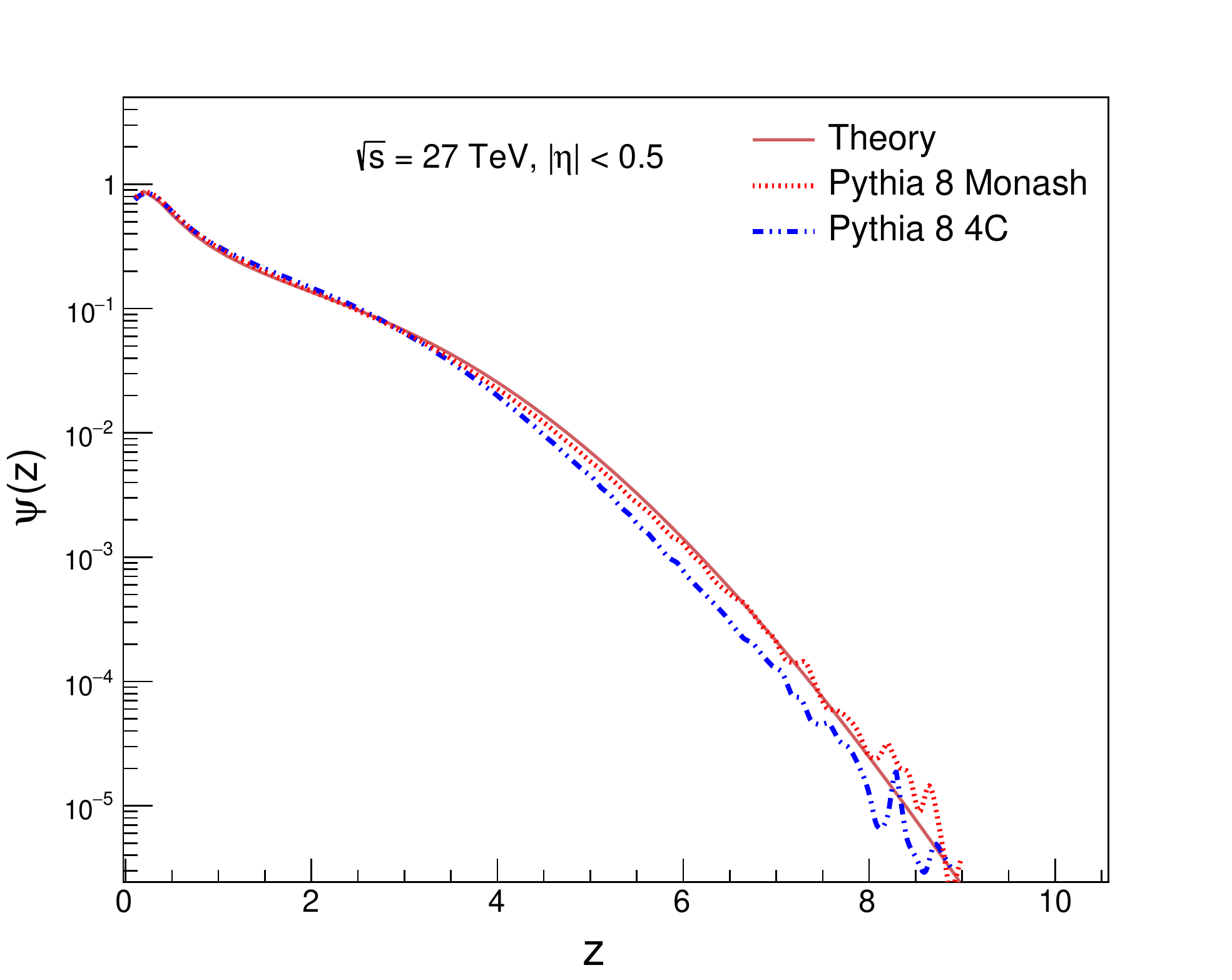}
\caption{KNO distributions at $\sqrt{s}$ = 13, 13.6 and 27 TeV in the
pseudorapidity interval $|\eta| < 0.5$.~Solid lines represent the predictions from the theoretical model and dotted lines represent the distributions generated from PYTHIA for two different tunes at each energy.}
\label{fig:KNO_05}
\end{figure}

\begin{figure}[htbp!]
\centering
\includegraphics[width=0.48\textwidth]{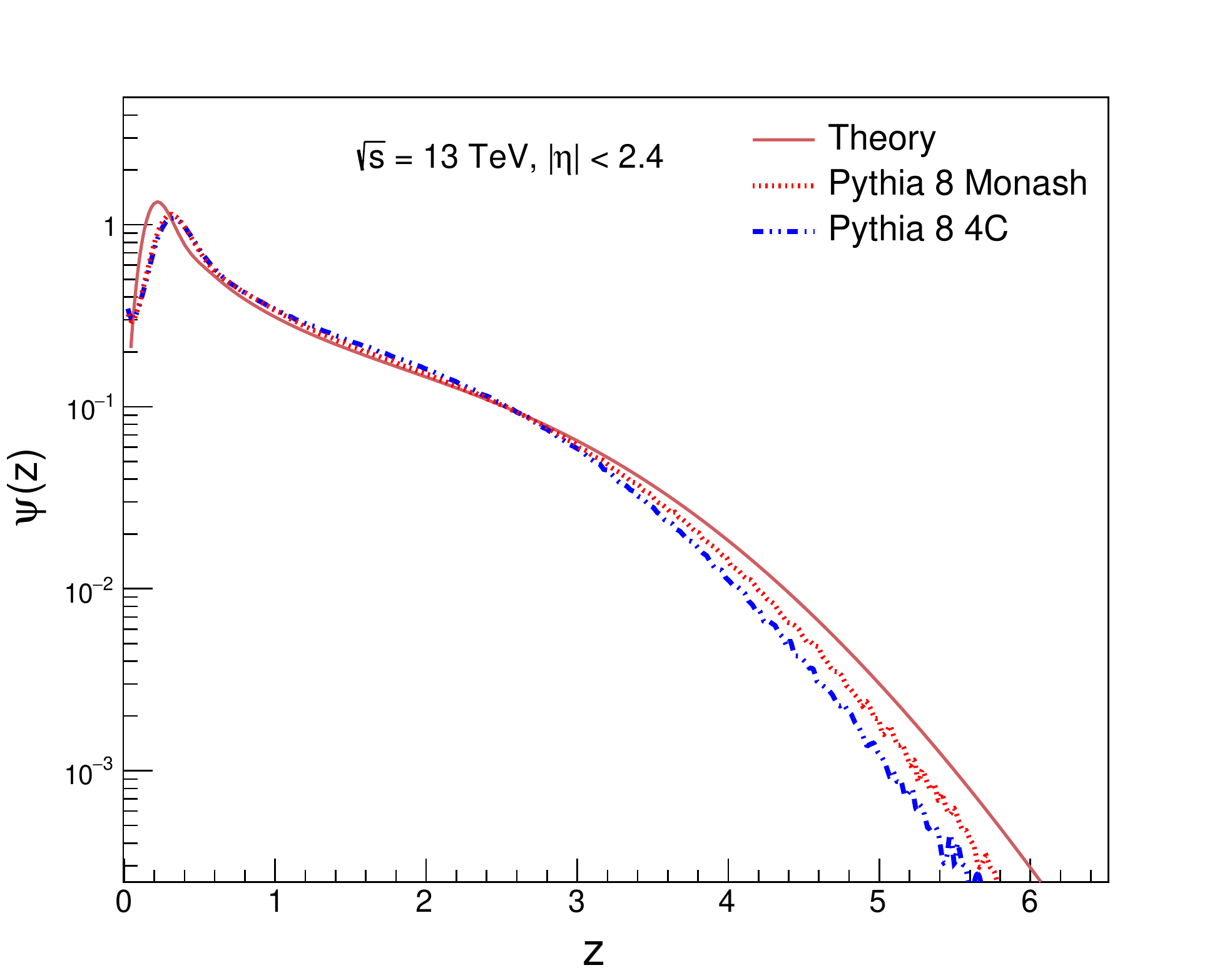}
\includegraphics[width=0.48\textwidth]{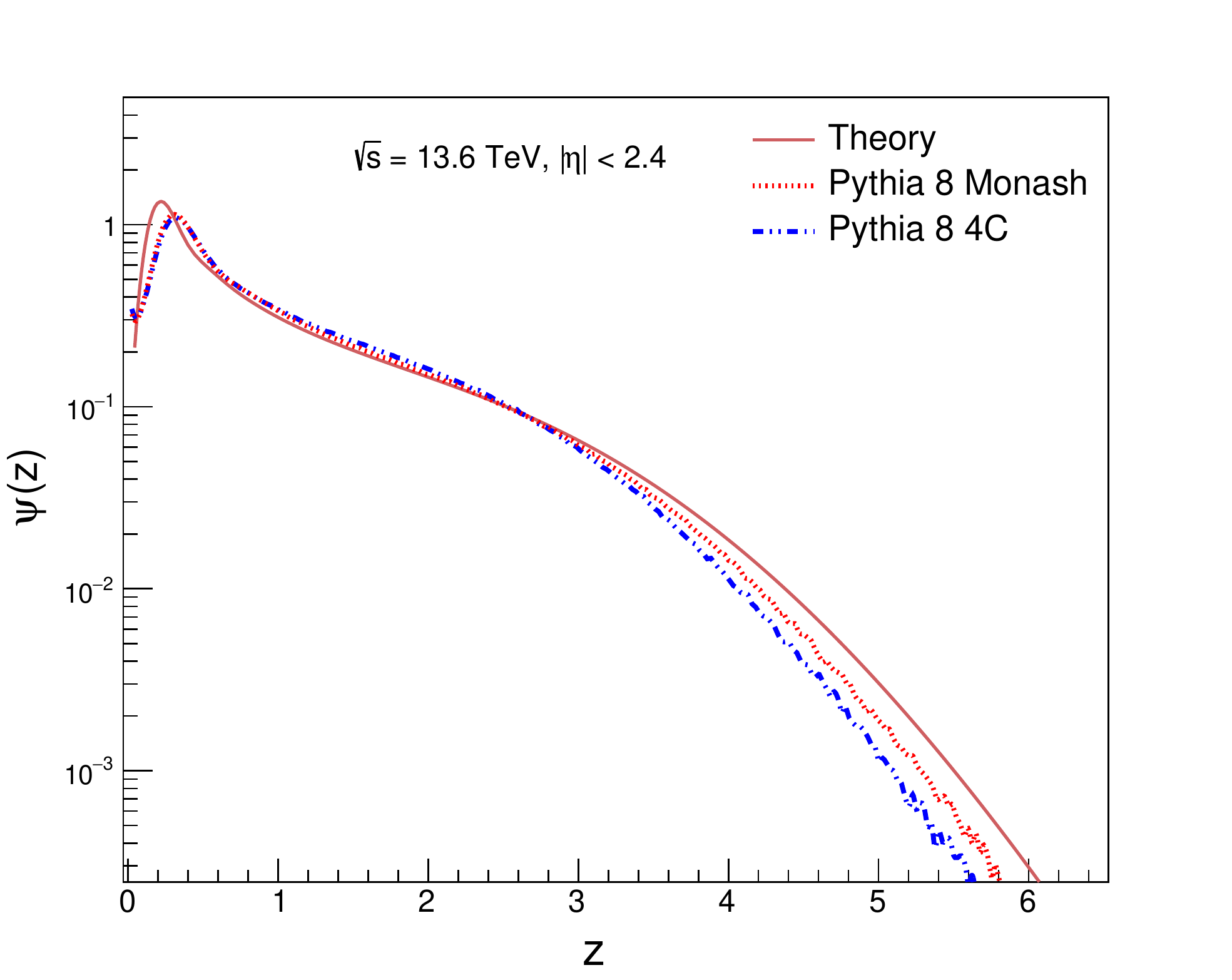}
\includegraphics[width=0.48\textwidth]{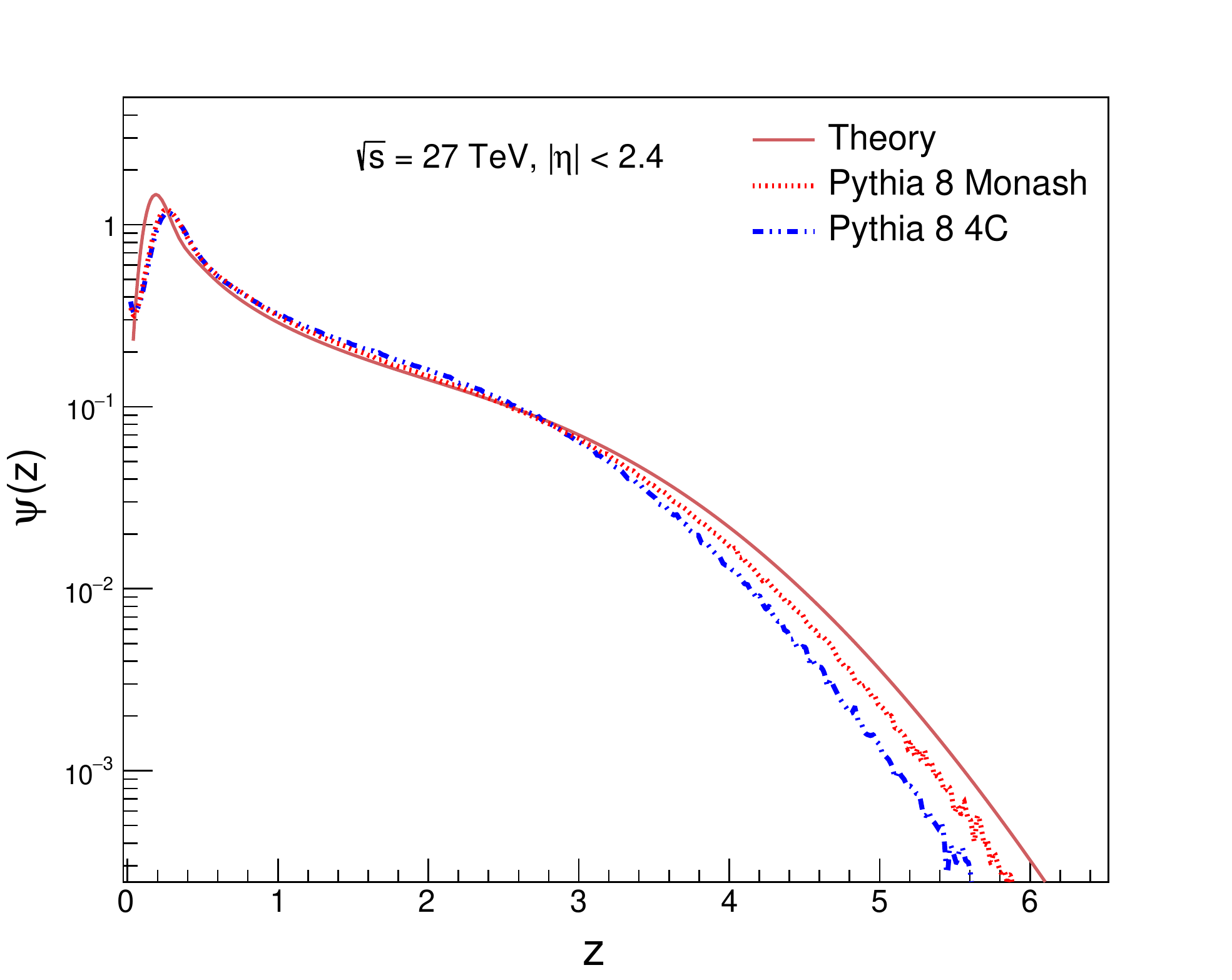}
\caption{KNO distributions at $\sqrt{s}$ = 13, 13.6 and 27~TeV in the pseudorapidity interval $|\eta| < 2.4$.~Solid lines represent the predictions from the theoretical model and dotted lines represent the distributions generated from PYTHIA for two different tunes at each energy.}
\label{fig:KNO_24}
\end{figure}

\begin{figure}[htbp!]
\centering
\includegraphics[width=0.48\textwidth]{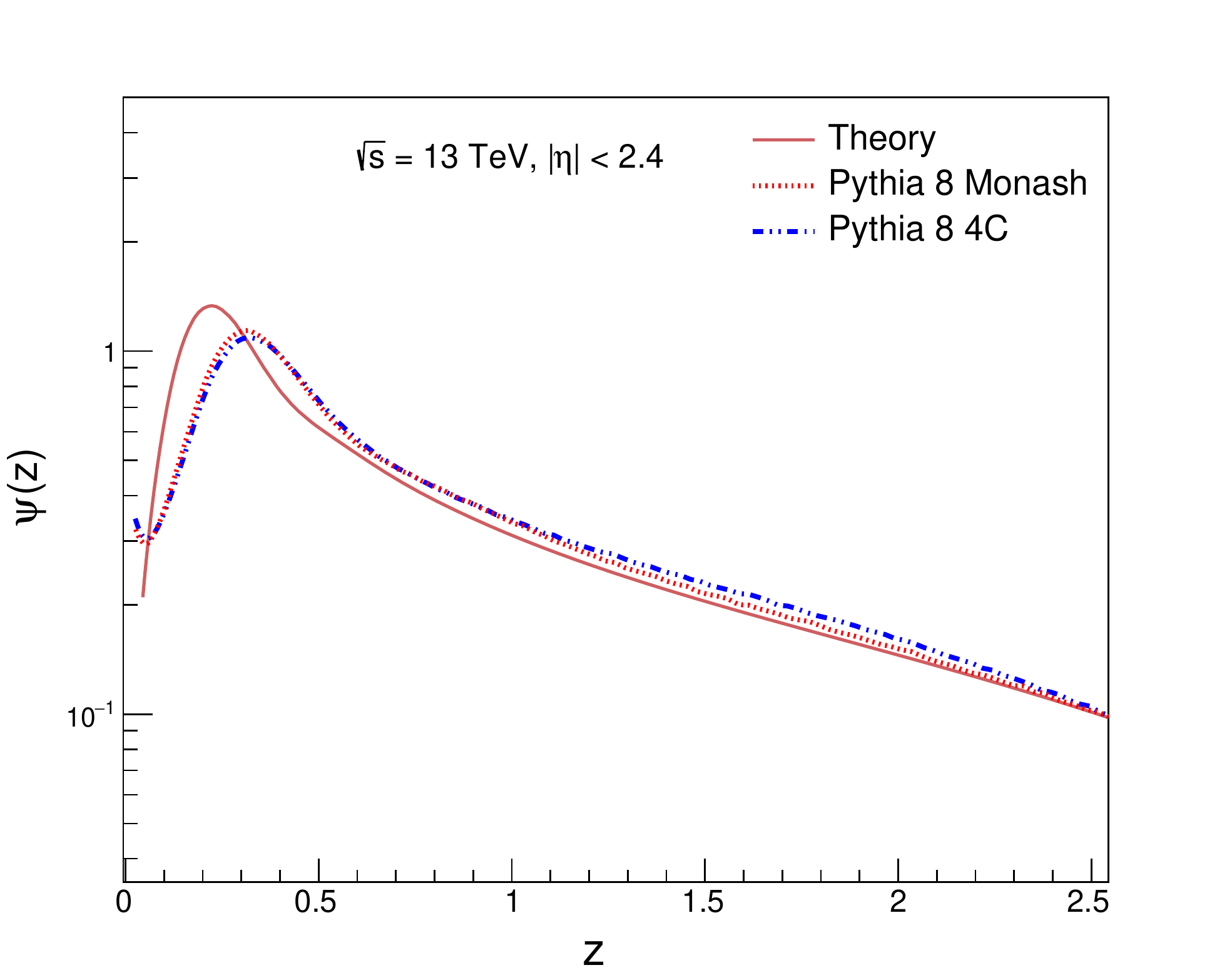}
\includegraphics[width=0.48\textwidth]{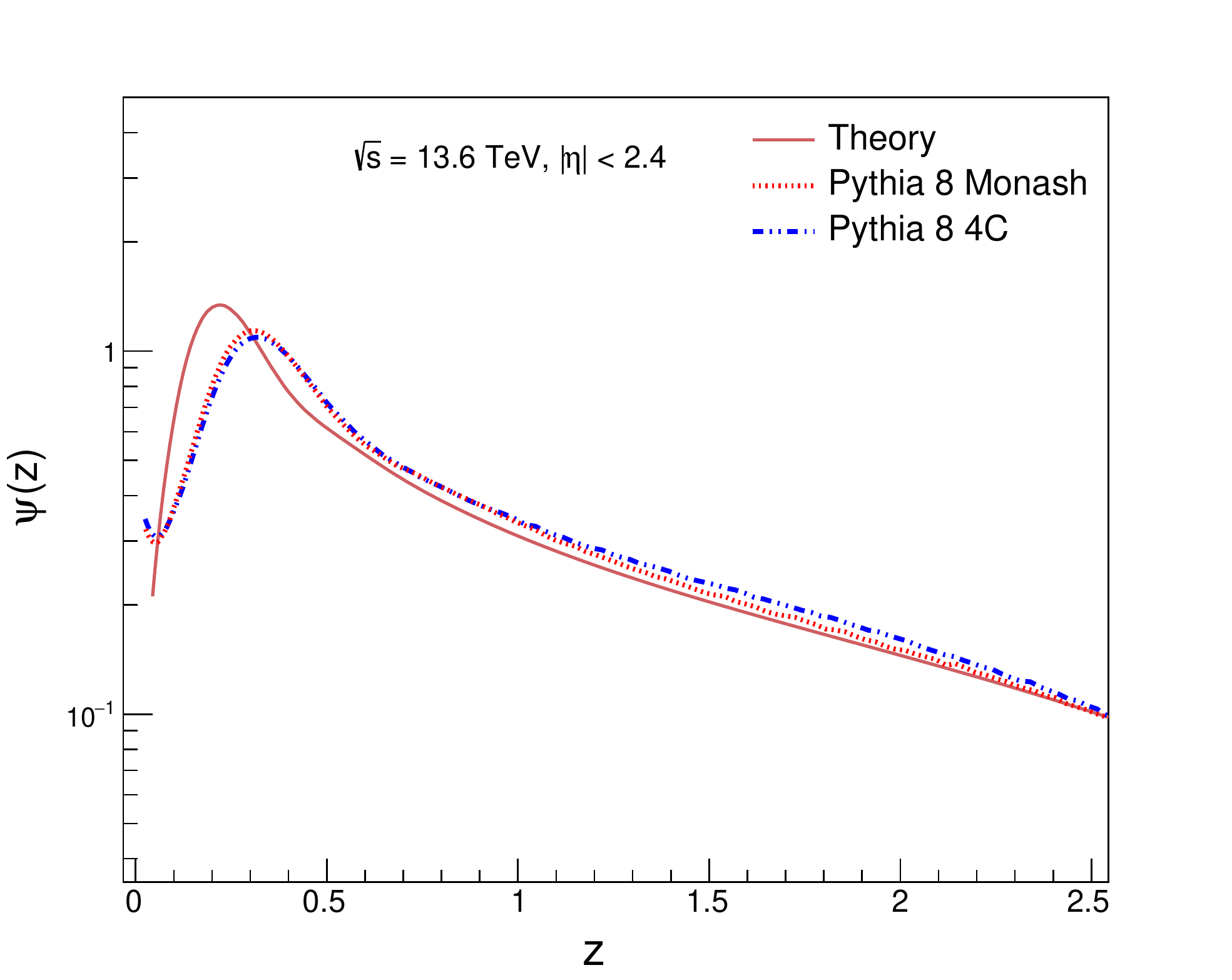}
\includegraphics[width=0.48\textwidth]{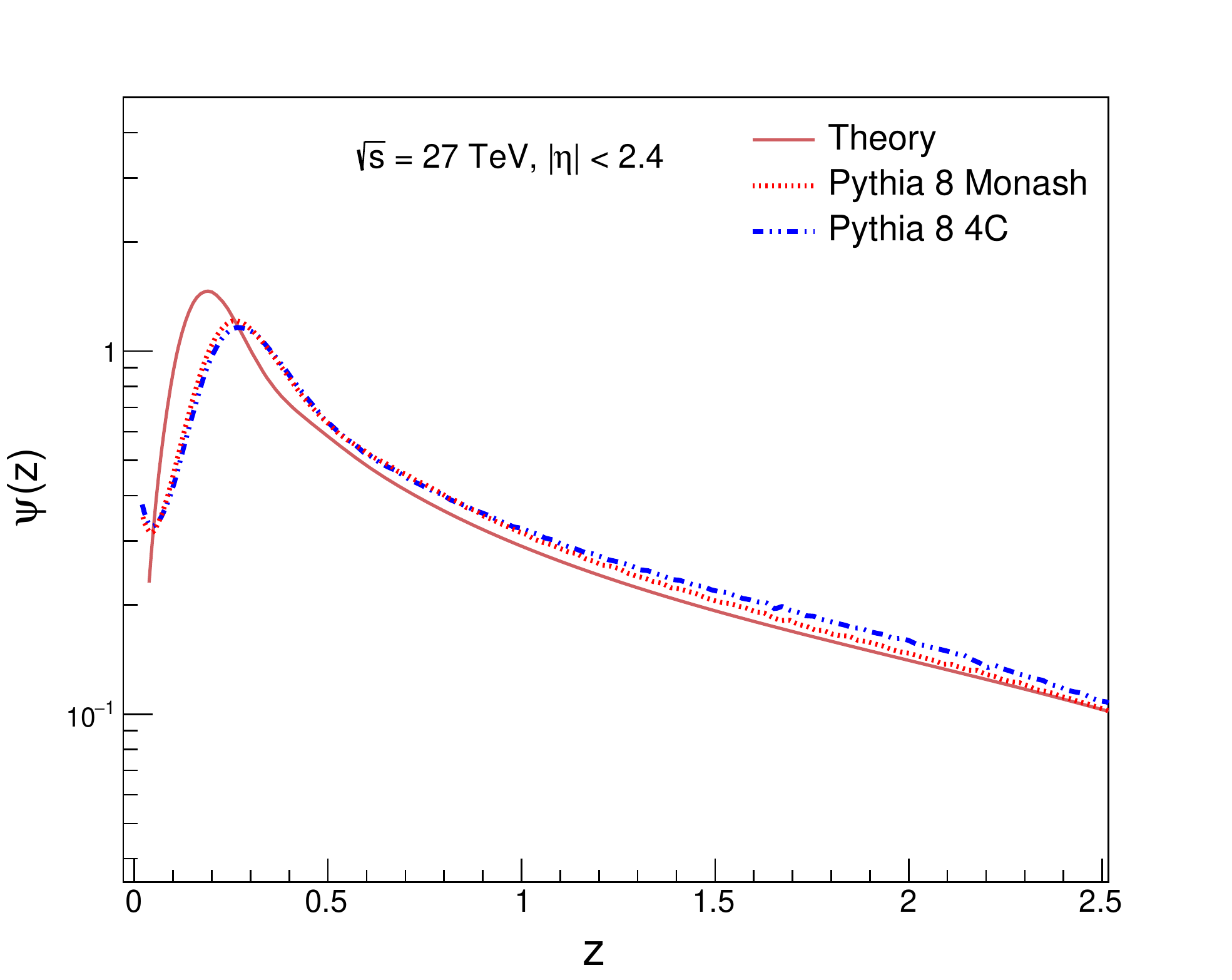}
\caption{~KNO distributions in the range $z$ = 0 to 2.5 for $\sqrt{s}$ = 13.6 and 27~TeV, in the pseudorapidity interval $|\eta| < $2.4.~Solid lines represent the predictions from the theoretical model and dotted lines represent the distributions generated from PYTHIA for two different tunes at each energy.}
\label{fig:peak2}
\end{figure}

The predictions for the KNO distributions from the DPM and PYTHIA 8 tunes are presented for the RUN2, RUN3 and future LHC energies at $\sqrt{s}$ = 13, 13.6 and 27~TeV in figures~\ref{fig:KNO_05} and \ref{fig:KNO_24} for $|\eta| <$ 0.5 and $|\eta| <$ 2.4 intervals respectively.~Also figure~\ref{fig:peak2} shows the KNO distributions around the peak values with $z <$ 2.5 at these energies in $|\eta| <$ 2.4 range.

It is also observed that the peak shifts towards smaller $z$ value as the collision energy increases from 0.9 TeV to 27 TeV. In the $|\eta| <$ 0.5 region, the predictions from DPM and PYTHIA 8 tunes agree, although PYTHIA Monash tune is closer to the theory.~For $|\eta| <$ 2.4, the two tunes of PYTHIA agree closely, however, the distribution from the DPM is shifted at each energy.~Below $z <$ 2.5, the model underestimates the PYTHIA 8 predictions, however, above $\sim$ 2.5 the model overestimates the PYTHIA 8 predictions.~Similar observations are made from Figure~\ref{fig:peak2} showing the distributions around the peak.

\section{CONCLUSION}
A detailed analysis and comparison of the charged hadron multiplicities in $pp$ collisions at various center of mass energies at the LHC is presented.~The analysis uses the data classified as non-single diffractive events obtained by the CMS experiment at $\sqrt{s}$ = 0.9, 2.36 and 7 TeV in the two pseudo-rapidity intervals, $|\eta| < 0.5$ and $|\eta| < 2.4$.~These data are compared with the  theoretical predictions of the Dual Parton Model and MC simulations of charged hadron production by using two different tunes of event generator PYTHIA 8.~Out of the two tunes used, Monash is the default tune in PYTHIA and 4C is the tune used by the CMS experiment.~Using these tunes and the theoretical calculations from the model, multiplicities are also obtained at $\sqrt{s}$ = 13, 13.6 and 27 TeV.~The LHC RUN3 has just started taking data at 13.6 TeV.~This analysis presents predictions for the charged hadron multiplicities at these energies and also for the future LHC energy of 27~TeV.

It is observed that the MDs in the $|\eta| < 0.5$ interval agree with the experimental distributions at $\sqrt{s}$ = 0.9, 2.36 and 7 TeV.~For the $|\eta| < 2.4$ interval, PYTHIA 8 predictions underestimate the experimental MDs in the mid multiplicity region for both Monash and 4C tunes, as seen in the ratio plot.~For the higher multiplicity region PYTHIA 8 predictions overestimate the experimental MDs.~In addition, a shoulder structure can be observed in the lowest multiplicity region.

The mean multiplicities obtained from the data, theory and MC are in good agreement for $\sqrt{s}$ = 0.9 and 2.36~TeV. The theoretical mean multiplicity soon starts to deviate from MC and the data at higher energy.~The $\langle{n}\rangle$ of the data and MC are in agreement within the error limit of experimental data at $\sqrt{s} = 7$~TeV.~However, theory systematically underestimates the experimental values from the CMS data, and PYTHIA overestimates.~The deviation of the theoretical mean gets more pronounced at centre of mass energy beyond 7~TeV.

KNO distributions obtained from the data~\citep{cms}, theory and MC are presented for $\sqrt{s}$ = 0.9, 2.36 and 7~TeV in Figures~\ref{fig:KNO_05} and~\ref{fig:KNO_24}.~For $|\eta| <$ 0.5 theoretical distributions are found to deviate from the data at $z \sim 6$, while for MC distributions the deviation is seen much earlier at $z \sim 4$.~Similarly, the KNO distributions at these energies in $|\eta| <$ 2.4 interval show deviation of theoretical distributions from the data at $z \sim 3$ and for MC distributions the deviation from the data starts at $z \sim 4$.~Exceptionally, the theoretical distribution from the model at $\sqrt{s}$ = 7~TeV follows the data very closely while the MC distributions show deviation.

The data from the CMS are not available for analysis at $\sqrt{s}$ = 13~TeV, for the NSD events in the same transverse momentum range.~The LHC RUN3 has started very recently and the data at the collision energy of 13.6~TeV is being collected by the experiments at the LHC.

We present the predictions for the mean multiplicities and the multiplicity distributions as estimated from the theoretical model and PYTHIA 8 for the two tunes for these energies.~Mean multiplicity and the KNO distributions for the future LHC energy of 27~TeV are also predicted.

It is also observed that the KNO distributions obtained from the two tunes agree very closely, though the low $z$ region below the peak shows disagreement between the theory and the data.~The peak of the KNO distribution in each case shifts towards smaller $z$ value as the collision energy increases from 0.9~TeV to 27~TeV.

The observations from the present study are indicative of the trends in the future data from the LHC.~Comparison with the data when it becomes available makes an interesting study and may lead to a new direction in our current understanding. 



\bibliography{bib.bib}

\end{document}